\documentclass[prd,twocolumn,superscriptaddress,altaffilletter,nofootinbib]{revtex4}

\usepackage{cancel}
\usepackage{graphicx}
\usepackage{amsmath}
\usepackage{amssymb}
\usepackage{graphicx,epsfig}
\newtheorem{theorem}{Theorem}
\newtheorem{corollary}{Corollary}[theorem]

\newcommand{\be}{\begin{equation}}
\newcommand{\ee}{\end{equation}}
\newcommand{\bea}{\begin{eqnarray}}
\newcommand{\eea}{\end{eqnarray}}
\newcommand{\der}{\partial}
\newcommand{\vphi}{\varphi}
\newcommand{\bet}{\begin{theorem}}
\newcommand{\eet}{\end{theorem}}
\newcommand{\bec}{\begin{corollary}}
\newcommand{\eec}{\end{corollary}}

\begin{document}



\title{Gauge invariant theory of gravity in spacetime with gradient nonmetricity:\\
A possible resolution of several cosmological puzzles}



\author{Israel Quiros}
\affiliation{Dpto. Ingenier\'ia Civil, Divisi\'on de Ingenier\'ia, Universidad de Guanajuato, Gto., M\'exico.}



\begin{abstract}
In this paper we apply the symmetry principle in order to search for an alternative unified explanation of several cosmological puzzles such as the present stage of accelerated expansion of the Universe and the Hubble tension issue, among others. We argue that Weyl gauge symmetry, being a manifest symmetry of gauge invariant theories of gravity operating on Weyl integrable geometry spacetimes, may be an actual (unbroken) symmetry of our present Universe. This symmetry may be at the core of a phenomenologically feasible explanation of modern fundamental issues arising within the framework of general relativity and of its known modifications.
\end{abstract}



\maketitle


\section{Introduction}

General relativity is the theory of gravity that has reigned for almost a century providing the correct description of gravitational phenomena within distance scales ranging from below a millimeter and up to cosmological scales. This theory has endured the onslaught of experimental and observational testing. However, at the end of the past and beginning of the present centuries the situation drastically changed. Experimental and observational evidence gathered which suggested that new physics is required. Several issues arose: dark matter (DM), dark energy (DE) and the cosmological constant problems among them, which pointed at new phenomenology. By relying on scalar field based models such as quintessence, k-essence, quintom, etc. many scientists and groups of scientists tried to find feasible explanation to these issues within the framework of general relativity (GR), without evident success. Others searched for modifications of GR: extra-dimensions, curvature modifications such as the $f(R)$ and teleparallel theories, etc. also without appreciable success. Quite the contrary, the increasing precision of experiments carries new experimental evidence that is uncovering new issues (for instance, the Hubble tension issue.) The present overall picture consists of a large zoo of diverse models which are designed to explain specific issues.

The situation reminds the state of affairs in the 60s of the past century in nuclear and in particle physics when there was plenty of theoretical models: different nuclear reactions were explained with the help of different models. Besides, a large amount of particles, quasi-particles and particle states were classified without a unified view. A clear way out of this situation was found with the guide of symmetries. The development of general relativity itself was tightly linked with the symmetry-based approach. We wonder, why not to apply the symmetry principle to search for resolution of modern fundamental issues such as the origin and nature of DM and of the DE? or, in other words: is there any missing symmetry which plays an important part in the classical description of the gravitational interactions of matter? 

In the bibliography one can find some interesting ideas which can help us in the search for an answer to the above question. In Ref. \cite{dicke-1962}, for instance, it is stated that the laws of physics must be invariant under coordinate-dependent transformations of units (conformal transformations) where the coordinate system is to be held fixed. In a similar fashion, more recently it has been argued that Weyl invariance should be viewed in the same manner as general coordinate invariance: all theories should respect Weyl invariance \cite{waldron, waldron-1, waldron-2}. Meanwhile, in \cite{thooft-2015} it is argued that small time and distance scales seem not to be related to large time and distance scales because we fail to understand the symmetry of the local conformal transformations. Is there any chance that local conformal symmetry could be the missing symmetry of the classical gravitational interactions? If so, could this symmetry explain (at least several of) the current cosmological puzzles? The main difficulty with any scenario where local conformal symmetry plays a role is that, even if the laws of gravity were Weyl invariant (same as local conformal invariant,) given that one is obliged to choose a specific gauge in order to be able to make calculations and to compare with observations, Weyl symmetry was not manifest in Nature. This is why, apparently, the universe is not local conformal invariant. 

Coincidentally, the birth of gauge symmetry, the one that played the leading role in the development of the standard model of particles (SMP) and of gauge field theories, took place within the framework of gravitational theories \cite{weyl-1917, many-weyl-book, london-1927, dirac-1973, utiyama-1973, adler-book, maeder-1978, smolin-1979, cheng-1988, cheng-arxiv, perlick-1991, drechsler-1999}. The birth of gauge symmetry went hand in hand with the advent of a new geometric framework: Weyl geometry. It represented a generalization of Riemann geometry that admitted variation of the length of timelike vectors during parallel transport, which is quantified by the nonmetricity vector or Weyl gauge vector. In this geometric framework, in addition to the curvature of background space which is originated from the metric components being spacetime functions, the nonmetricity vector played an important role in the definition of the affine properties of space. The additional structure carried a new symmetry: invariance under Weyl rescalings, consisting of conformal transformation of the metric simultaneously with gauge transformations of the nonmetricity vector. This early symmetry approach to gravitation was unsuccessful since it was intended as a possible unification scheme of the electromagnetic and gravitational interactions. 

The possibility that Weyl symmetry may be an underlying symmetry in Nature has been explored in \cite{waldron, waldron-1, waldron-2}. In these bibliographic references the authors construct theories that unify massless, massive, and partially massless excitations. The method used by the authors relies on tractor calculus -- mathematical machinery allowing Weyl invariance to be kept manifest at all stages, thus challenging the common wisdom. The scope of the present paper is much more modest: here we shall not develop any new mathematics nor even any new physics. Our goal is to take a new look at well-known ``old'' mathematics and physics. We argue that if take a different look at local conformal symmetry, perhaps this symmetry may help us to understand those aspects of the gravitational interactions that could be missing and, for which reason there are several cosmological puzzles that have not been solved yet. 

In this paper we shall consider Weyl integrable geometry (WIG) spaces which are a particular case of Weyl geometry space \cite{weyl-1917, many-weyl-book, london-1927, dirac-1973, utiyama-1973, adler-book, maeder-1978}. The latter belongs, in turn, in the so called generalized Weyl geometry spacetimes, denoted here by $W_4$. Generalized Weyl geometry spaces are characterized by the distinctive property that the covariant derivative of the metric does not vanish \cite{delhom-2019}:

\bea \nabla_\alpha g_{\mu\nu}=-Q_{\alpha\mu\nu},\label{gen-nm}\eea where $Q_{\alpha\mu\nu}$ is the nonmetricity tensor.\footnote{Recently generalized nonmetricity theories, which are based in generalized Weyl geometry spacetimes, have played an important role in the search for alternative explanations to fundamental questions of current interest. The recent interest in nonmetricity theories is mainly focused in the so called teleparallel \cite{tele-rev-capoz, tele-ferraro, tele-maluf, tele-soti, tele-brazil, tele-baha, tele-nester, tele-coley, tele-intro} and, specially, the symmetric teleparallel theories \cite{adak-2006, adak-2006-1, adak-2013, beltran-plb-2016, javr-prd-2018, vilson-prd-2018, formiga-2019, adak-arxiv, obukhov, beltran-j-universe} and their cosmological applications \cite{lavinia-review, beltran-prd-2020, saridakis-prd-2020, lazkoz-prd-2019, lazkoz-prd-2021, sanjay-prd-2020, sanjay-2021}, under the implicit assumption that conformal symmetry plays no role. Here we shall not consider the teleparallel condition so that the mentioned works fall outside of the scope of the present paper.} Arbitrary nonmetricity brings with it several issues of fundamental character such as ambiguity in the definition of the gauge covariant derivative operators \cite{quiros-2022} and in the determination of the actual role geodesics and autoparallels play within the geometrical structure of generalized Weyl spaces \cite{adak-arxiv, obukhov} (in spaces with arbitrary nonmetricity the autoparallels and the geodesics do not coincide.) Vectorial nonmetricity, where $Q_{\alpha\mu\nu}=Q_\alpha g_{\mu\nu}$ ($Q_\mu$ is the Weyl gauge vector) and gradient nonmetricity, where $Q_{\alpha\mu\nu}=\der_\alpha\vphi g_{\mu\nu}$ ($\vphi$ is a scalar field), are free of the mentioned ambiguity. The phenomenological consequences of gauge invariant theories of gravity with vectorial nonmetricity have been explored in \cite{quiros-arxiv-2022}, where it has been demonstrated that these are not able to take account of the dynamics of the Universe beyond the radiation-dominated stage of the cosmic expansion. In the present paper, in order to complement the former study, we shall investigate the physical implications of gauge invariant theories of gravity with gradient nonmetricity. As it was shown in \cite{quiros-arxiv-2022}, this is the only possibility left to us by Nature for gauge symmetry to have an impact in the cosmic dynamics beyond radiation domination.  

 
Local conformal symmetry is one of the fundamental properties of spaces with nonmetricity. It amounts to invariance under Weyl rescalings not only of the gravitational action -- as well as of the derived equations of motion (EOM) -- but also of the geometrical structure: the nonmetricity law, the autoparallels and geodesics, among other relevant geometric equations and quantities. Weyl rescalings or local conformal transformations consist of conformal transformations of the metric $g_{\mu\nu}$ and simultaneous gauge transformation of the nonmetricity tensor $Q_{\alpha\mu\nu}$, plus transformations of the remaining fields $\Psi$ according to their conformal weight $w$. 

For spaces with vectorial nonmetricity $Q_\mu$ the Weyl rescalings read,

\bea g_{\mu\nu}\rightarrow\Omega^2g_{\mu\nu},\;Q_\alpha\rightarrow Q_\alpha-2\der_\alpha\ln\Omega,\;\Psi\rightarrow\Omega^w\Psi,\label{gauge-t}\eea where the positive smooth function $\Omega$ is the conformal factor. In this paper, for brevity and because it is historically justified, we shall call the transformations \eqref{gauge-t} as gauge transformations and the associated symmetry as gauge symmetry. Spaces with vectorial nonmetricity are also called as Weyl spaces and are denoted here by $\tilde W_4$. Obviously Weyl space is a subspace of generalized Weyl space: $\tilde W_4\subset W_4$. The nonmetricity law for Weyl space is given by

\bea \nabla_\alpha g_{\mu\nu}=-Q_{\alpha} g_{\mu\nu},\label{vect-nm}\eea where the covariant derivative operator $\nabla_\mu$ is defined in terms of the affine connection of $\tilde W_4$ space ($\Gamma^\alpha_{\;\;\mu\nu}$), which can be decomposed in terms of the Levi-Civita affine connection of Riemann space $V_4$ (Christoffel symbols of the metric) $\{^\alpha_{\mu\nu}\}$ and of the disformation tensor $L^\alpha_{\;\;\mu\nu}$, in the following way:

\bea \Gamma^\alpha_{\;\;\mu\nu}=\{^\alpha_{\mu\nu}\}+L^\alpha_{\;\;\mu\nu},\label{gen-aff-c}\eea with 

\bea &&\{^\alpha_{\mu\nu}\}:=\frac{1}{2}g^{\alpha\lambda}\left(\der_\nu g_{\mu\lambda}+\der_\mu g_{\nu\lambda}-\der_\lambda g_{\mu\nu}\right),\label{lc-aff-c}\\
&&L^\alpha_{\;\;\mu\nu}:=\frac{1}{2}\left(Q_\mu\delta^\alpha_\nu+Q_\nu\delta^\alpha_\mu-Q^\alpha g_{\mu\nu}\right).\label{disf-t}\eea The Weyl gauge vector $Q_\alpha$ measures how much the length of given timelike vector varies during parallel transport. 

In this paper the curvature tensor of $\tilde W_4$ spacetime, or properly, the curvature of the connection, is defined as it follows,

\bea &&R^\alpha_{\;\;\sigma\mu\nu}:=\der_\mu\Gamma^\alpha_{\;\;\nu\sigma}-\der_\nu\Gamma^\alpha_{\;\;\mu\sigma}\nonumber\\
&&\;\;\;\;\;\;\;\;\;\;\;\;\;\;\;+\Gamma^\alpha_{\;\;\mu\lambda}\Gamma^\lambda_{\;\;\nu\sigma}-\Gamma^\alpha_{\;\;\nu\lambda}\Gamma^\lambda_{\;\;\mu\sigma},\label{gen-curv-t}\eea or, if take into account the decomposition \eqref{gen-aff-c}:

\bea &&R^\alpha_{\;\;\sigma\mu\nu}=\hat R^\alpha_{\;\;\sigma\mu\nu}+\hat\nabla_\mu L^\alpha_{\;\;\nu\sigma}-\hat\nabla_\nu L^\alpha_{\;\;\mu\sigma}\nonumber\\
&&\;\;\;\;\;\;\;\;\;\;\;\;\;\;+L^\alpha_{\;\;\mu\lambda}L^\lambda_{\;\;\nu\sigma}-L^\alpha_{\;\;\nu\lambda}L^\lambda_{\;\;\mu\sigma},\label{gen-curv-t-1}\eea where $\hat R^\alpha_{\;\;\sigma\mu\nu}$ is the Riemann-Christoffel or LC curvature tensor,

\bea &&\hat R^\alpha_{\;\;\sigma\mu\nu}:=\der_\mu\{^\alpha_{\nu\sigma}\}-\der_\nu\{^\alpha_{\mu\sigma}\}\nonumber\\
&&\;\;\;\;\;\;\;\;\;\;\;\;\;\;\;+\{^\alpha_{\mu\lambda}\}\{^\lambda_{\nu\sigma}\}-\{^\alpha_{\nu\lambda}\}\{^\lambda_{\mu\sigma}\},\label{lc-curv-t}\eea and $\hat\nabla_\alpha$ is the LC covariant derivative.\footnote{In this paper a hat over a quantity or operator means that it is defined with respect to the LC affine connection \eqref{lc-aff-c}, i. e., that it is a Riemannian quantity or operator.} Besides, the LC Ricci tensor $\hat R_{\mu\nu}=\hat R^\lambda_{\;\;\mu\lambda\nu}$ and LC curvature scalar read:

\bea &&\hat R_{\mu\nu}=\der_\lambda\{^\lambda_{\nu\mu}\}-\der_\nu\{^\lambda_{\lambda\mu}\}+\{^\lambda_{\lambda\kappa}\}\{^\kappa_{\nu\mu}\}-\{^\lambda_{\nu\kappa}\}\{^\kappa_{\lambda\mu}\},\nonumber\\
&&\hat R=g^{\mu\nu}\hat R_{\mu\nu},\label{lc-curv-sc}\eea respectively. It can be shown that the curvature scalar of $\tilde W_4$ can be decomposed in the following way:

\bea R=\hat R-\frac{3}{2}Q_\mu Q^\mu-3\hat\nabla_\mu Q^\mu.\label{curv-decomp}\eea 


In a recent publication \cite{quiros-arxiv-2022} we have shown that the bi-parametric class of gravitational Lagrangians over $\tilde W_4$ space,

\bea {\cal L}_g=\alpha_1\sqrt{-g}\left[\phi^2R+\omega(\der^*\phi)^2-\frac{\beta^2}{2}Q_{\mu\nu}Q^{\mu\nu}\right],\label{g-lag}\eea where $\omega$ and $\beta$ are free constants,\footnote{The overall constant factor $\alpha_1$ affects only the matter source of the Einstein's like equations so that it may be absorbed into the measured Newton's constant.} $Q_{\mu\nu}\equiv2\der_{[\mu}Q_{\nu]}$ and we use the notation $(\der^*\phi)^2\equiv g^{\mu\nu}\der^*_\mu\phi\der^*_\nu\phi$, with the gauge derivative given by\footnote{Let ${\bf T}$ be a $(p,q)$-tensor in $\tilde W_4$, with coordinate components $T^{\alpha_1\alpha_2\cdots\alpha_p}_{\beta_1\beta_2\cdots\beta_q}$ and with conformal weight $w({\bf T})=w$, so that under \eqref{gauge-t}: ${\bf T}\rightarrow\Omega^w{\bf T}$. Then, the Weyl gauge derivative of the tensor reads: $$\der^*_\alpha{\bf T}:=\der_\alpha{\bf T}+\frac{w}{2}Q_\alpha{\bf T}.$$ This definition warrants that the gauge derivative transforms like the tensor itself, i. e., under \eqref{gauge-t}: $\der^*_\alpha{\bf T}\rightarrow\Omega^w\der^*_\alpha{\bf T}$.} $\der^*_\mu\phi=(\der_\mu-Q_\mu/2)\phi$, are the mathematical basis of gauge invariant theories of gravity where only radiation and massless fields with vanishing trace of the stress-energy tensor (SET) couple to gravity. Hence, the phenomenology after $SU(2)\times U(1)$ symmetry breaking falls beyond the scope of these theories. In other words, fields of the SMP with nonvanishing mass do not interact with gravity in this class of theories, so that they can have an impact in the description of the phenomenology, at most, during the radiation-dominated stage of the cosmic expansion. This means, in particular, that the class of gauge invariant theories of gravity given by the Lagrangian \eqref{g-lag} can not explain the matter-dominated epoch of the cosmic evolution, including the dark ages: DM and DE dominated stages. 


The above result is valid as well for gauge invariant theories of gravity which are based in the traceless Weyl tensor. For instance, in \cite{mannheim-1989, mannheim-2006} a conformal invariant quadratic theory of gravity has been proposed, which operates in Riemann space $V_4$. This gauge invariant gravitational theory has been designed to solve several problems, including a possible explanation to the dark matter issue. The proposed gauge invariant Lagrangian reads,\footnote{This theory has many problems. For instance, it does not have the Einstein-Hilbert (low-curvature) limit, since the gravitational spectrum contains only the ghost-like spin-two field and has no graviton in it. This rules out this theory as a phenomenologically viable description of low curvature gravitational phenomena.}

\bea {\cal L}_{C^2}=\alpha\sqrt{-g}\hat C^2=2\alpha\sqrt{-g}\left[\hat R_{\mu\nu}\hat R^{\mu\nu}-\frac{1}{3}\hat R^2\right],\label{mannheim-action}\eea where $\alpha$ is a dimensionless constant, $\hat C^2\equiv\hat C_{\mu\lambda\nu\sigma}\hat C^{\mu\lambda\nu\sigma}$, $\hat C_{\mu\lambda\nu\sigma}$ is the Weyl tensor of Riemann space and we have used the Gauss-Bonnet invariant to trade the $\hat C^2$-term by the terms within square brackets in \eqref{mannheim-action}. The derived EOM reads \cite{mannheim-1989}:

\bea \hat W^{(2)}_{\mu\nu}-\frac{1}{3}\hat W^{(1)}_{\mu\nu}=\frac{1}{4\alpha}T^\text{mat}_{\mu\nu},\label{mannheim-eom}\eea where 

\bea &&\hat W^{(1)}_{\mu\nu}=2g_{\mu\nu}\hat\nabla^2\hat R-2\hat\nabla_\mu\hat\nabla_\nu\hat R\nonumber\\
&&\;\;\;\;\;\;\;\;\;\;\;\;-2\hat R\hat R_{\mu\nu}+\frac{1}{2}g_{\mu\nu}\hat R^2,\nonumber\\
&&\hat W^{(2)}_{\mu\nu}=\frac{1}{2}g_{\mu\nu}\hat\nabla^2\hat R+\hat\nabla^2\hat R_{\mu\nu}-2\hat\nabla_{(\mu}\hat\nabla_\lambda\hat R^{\;\;\lambda}_{\nu)}\nonumber\\
&&\;\;\;\;\;\;\;\;\;\;\;\;-2\hat R_{\mu\lambda}\hat R^{\;\;\lambda}_\nu+\frac{1}{2}g_{\mu\nu}\hat R_{\lambda\sigma}\hat R^{\lambda\sigma}.\nonumber\eea As shown in \cite{quiros-arxiv-2022}, the trace of Eq. \eqref{mannheim-eom} reads,

\bea \frac{1}{4\alpha}T^\text{mat}=\hat\nabla^2\hat R-2\hat\nabla^\mu\hat\nabla^\nu\hat R_{\mu\nu}=-2\hat\nabla^\mu\hat\nabla^\nu\hat G_{\mu\nu},\nonumber\eea which exactly vanishes thanks to the Bianchy identity $\hat\nabla^\mu\hat G_{\mu\nu}=0$. Hence, only radiation couples to gravity in this theory, so that quite the contrary effect is obtained since the DM does not interact with radiation.


In \cite{quiros-arxiv-2022} it was concluded that, since on the one hand only massless fields, which do not interact with nonmetricity, can be coupled to gravity in a theory based in the class of gauge invariant gravitational Lagrangians \eqref{g-lag} -- also valid for \eqref{mannheim-action} -- and on the other hand, after breakdown of electroweak (EW) symmetry and acquirement of mass by the SMP fields, gauge symmetry should be a broken symmetry, then there is no place for vectorial nonmetricity, and for the associated gauge symmetry, in the classical description of the gravitational interactions of matter. Instead, gradient nonmetricity and its associated gauge symmetry, seem to have an opportunity to play a role in the gravitational interactions of matter at the classical level. Gradient nonmetricity: $\nabla_\alpha g_{\mu\nu}=-\der_\alpha\vphi g_{\mu\nu}$, where $\vphi$ is a scalar function, is the defining property of WIG spaces, which we shall denote here by $\tilde W^\text{int}_4$. It is obvious that the following relationship is satisfied: $\tilde W^\text{int}_4\subset\tilde W_4\subset W_4$.

The assumption that WIG may be the geometrical substrate for gauge invariant (classical) gravitational interactions of matter entails a serious challenge. If a gauge invariant gravitational theory over $\tilde W^\text{int}_4$ space is capable of correctly describing the classical gravitational interactions of matter, then fields with mass, which follow timelike geodesics of WIG space, must couple to Weyl integrable geometry, i. e., any timelike matter fields must couple both to the curvature of space and to gradient nonmetricity\footnote{This is not true for massless fields, which always follow null geodesics of Riemann space.} in such a way that gauge symmetry is preserved. This means that gauge symmetry must survive $SU(2)\times U(1)$ symmetry breaking. The question is: can EW symmetry breaking and gauge symmetry coexist together? 

EW symmetry breaking may be associated with the Higgs Lagrangian,

\bea {\cal L}_H=-\frac{\sqrt{-g}}{2}\left[|D_gH|^2+\lambda'\left(|H|^2-v^2_0\right)^2\right],\label{higgs-lag}\eea where $v_0$ is the EW mass parameter, $H$ is the Higgs isodoublet, and we use the following notation: $|H|^2\equiv H^\dag H$, $|D_gH|^2\equiv g^{\mu\nu}(D^g_\mu H)^\dag(D^g_\nu H)$,

\bea D^g_\mu H=\left(\der_\mu+\frac{i}{2}g W^k_\mu\sigma^k+\frac{i}{2}g' B_\mu\right)H,\label{ew-gauge-der}\eea with $W^k_\mu$ -- the $SU(2)$ bosons, $B_\mu$ -- the $U(1)$ boson, $(g,g')$ -- gauge couplings and $\sigma^k$ are the Pauli matrices. Under the conformal transformation of the metric, 

\bea &&(H,H^\dag)\rightarrow\Omega^{-1}(H,H^\dag),\;v_0\rightarrow v_0,\nonumber\\
&&D^g_\mu H\rightarrow\Omega^{-1}\left(D^g_\mu-\der_\mu\ln\Omega\right)H,\nonumber\\
&&\left(D^g_\mu H\right)^\dag\rightarrow\Omega^{-1}\left[\left(D^g_\mu H\right)^\dag-\der_\mu\ln\Omega H^\dag\right],\nonumber\eea so that the Higgs Lagrangian ${\cal L}_H$, is not invariant under the gauge transformations \eqref{gauge-t}. Hence, if we expect gauge symmetry to survive EW symmetry breaking, the Lagrangian \eqref{higgs-lag} has to be modified. 

The required modification amounts to lifting the mass parameter $v_0$ to a point dependent field \cite{bars-2014}: $v_0\rightarrow v({\bf x})$,  with conformal weight $w=-1$, such that under \eqref{gauge-t}, $v^2\rightarrow\Omega^{-2}v^2$. Besides, the EW gauge covariant derivative in Eq. \eqref{ew-gauge-der} is to be replaced as well: $D^g_\mu\rightarrow D^{*g}_\mu\equiv D^g_\mu-Q_\mu/2$, so that, under the gauge transformations \eqref{gauge-t}, $D^{*g}_\mu H\rightarrow\Omega^{-1}D^{*g}_\mu H$, $(D^{*g}_\mu H)^\dag\rightarrow\Omega^{-1}(D^{*g}_\mu H)^\dag$ $\Rightarrow\;|D^*_gH|^2\rightarrow\Omega^{-4}|D^*_gH|^2$. Lifting of the mass parameter to a point dependent field $v({\bf x})$ leads to the masses acquired by the particles of the SMP after EW symmetry breaking, being point dependent quantities as well: $m=m({\bf x})$. Hence, under \eqref{gauge-t} the mass $m$ of given particle transforms like $m\rightarrow\Omega^{-1}m$. Means that Weyl gauge symmetry may survive after $SU(2)\times U(1)$ symmetry breaking and thus it may play a role in the past, present and future of the cosmic evolution of our Universe. Notice that this approach is very different from the one undertaken in the bulk of papers on gauge theories of gravity, where the Weyl gauge symmetry breaks down either through the Higgs procedure or through other alternative mechanisms \cite{ghilen-1, ghilen-2, ghilen-3, bars-2014}. 

In order to illustrate the consequences of the modification of the Higgs Lagrangian \eqref{higgs-lag} proposed in \cite{bars-2014}, let us discuss on the following well-known example. In Ref. \cite{deser-1970} in order to break the gauge symmetry of the gravitational Lagrangian 

\bea {\cal L}_g=\sqrt{-g}\left[\frac{\phi^2}{12}\hat R+\frac{1}{2}(\der\phi)^2\right],\label{deser-lag}\eea a mass term for the scalar field $\phi$,

\bea {\cal L}_\mu=\frac{\sqrt{-g}}{2}\mu^2\phi^2,\label{sb-lag}\eea is added. Then, under the assumption that the constant mass $\mu$ of the scalar field is not transformed by \eqref{gauge-t}: $\mu\rightarrow\mu$, the Lagrangian ${\cal L}_\mu\rightarrow\Omega^2{\cal L}_\mu$, so that gauge invariance is spoiled. Of course, this conclusion is based on the Higgs Lagrangian \eqref{higgs-lag}, which leads to the fields $\psi$ of the SMP acquiring constant masses $m_\psi\propto g_\psi v_0$, where $g_\psi$ is the Yukawa coupling for the field $\psi$. However, a drastically different result is obtained if, following the above discussed modification of the Higgs Lagrangian, in Eq. \eqref{higgs-lag} replace the mass parameter by a point-dependent field $v_0\rightarrow v({\bf x})$. In this case the mass acquired by the scalar field would be also a point dependent quantity $\mu\rightarrow\mu({\bf x})\propto v({\bf x})$ so that, under the gauge transformations \eqref{gauge-t}, $\mu\rightarrow\Omega^{-1}\mu$ $\Rightarrow{\cal L}_\mu\rightarrow{\cal L}_\mu$. I. e. the gauge symmetry is preserved by the Lagrangian \eqref{sb-lag}. This teaches us that we should be open minded in regard to gauge symmetry and the inclusion of masses since, as seen, there is a possibility that masses $m$ transform like $m\rightarrow\Omega^{-1}m$ under the gauge transformations \eqref{gauge-t}, as in Ref. \cite{dicke-1962} and in works on scalar-tensor theories, where under a conformal transformation of the metric the masses of particles transform precisely in this way \cite{faraoni-rev, faraoni-prd-2007, quiros-rev-2019}.

Weyl integrable geometry is usually underestimated by identifying it with Riemann geometry, while the related theory of gravity is incorrectly identified with general relativity.\footnote{There are, however, several works \cite{novello-1992, romero-1, romero-2, romero-3, romero-4} where the role played by WIG in the description of the gravitational phenomena is investigated.} As a matter of fact WIG is a class of geometries while the related gauge invariant theory of gravity is a class of theories. GR is just a specific gauge in this class. Different gauges represent different theories (see the related discussion in \cite{quiros-arxiv-2022} and in sections \ref{sect-gauges} and \ref{sect-many-w} of this paper.) As we shall see these can be differentiated through the check of the observational and experimental evidence (see sections \ref{sect-z-q} and \ref{sect-observ} below.) 

We want to notice that there are clear differences between gauge invariance within a gravitational theory and, for instance, electromagnetic (EM) $U(1)$ gauge invariance. In the latter case Maxwell's and Dirac's equations are invariant under the $U(1)$ transformations:

\bea &&\psi\rightarrow e^{-ie\lambda(x)}\psi,\;\bar\psi\rightarrow e^{ie\lambda(x)}\bar\psi,\nonumber\\
&&A_\mu\rightarrow A_\mu+\der_\mu\lambda(x),\label{u1-t}\eea where $A_\mu$ is the EM vector potential, $\psi$ is the fermion's spinor and $\lambda(x)$ can be any function. Any two states, picked out by two different choices $\lambda_1(x)$ and $\lambda_2(x)$, are to be identified. This is due to the fact that the probability density $\propto\bar\psi\psi$, which carries the relevant information about the quantum state of the fermion, is not affected by phase shifts $\sim\lambda(x)$, i. e. the probability density $\bar\psi\psi\rightarrow\bar\psi\psi$ is invariant under \eqref{u1-t}. 

In the case of a gauge invariant theory of gravity in WIG space $\tilde W^\text{int}_4$ (gradient nonmetricity $Q_\mu=2\der_\mu\ln\phi$,) gauge invariance means that the gravitational equations of motion together with the EOM of the remaining matter fields, are not affected by the gauge transformations which are composed of conformal transformations of the metric $g_{\mu\nu}\rightarrow\Omega^2g_{\mu\nu}$, together with simultaneous gauge transformation of the geometric scalar $Q_\mu\rightarrow Q_\mu-2\der_\mu\ln\Omega$ $\Rightarrow\phi\rightarrow\Omega^{-1}\phi$, and appropriate transformations of the remaining fields. Contrary to $U(1)$ gauge symmetry in electrodynamics, there is nothing similar to probability density in the gauge invariant theory of gravity in $\tilde W^\text{int}_4$ space. Instead, the conformal transformation of the metric links two different metrics, i. e., two different ways of measuring distances in spacetime. Each one of the conformally related metrics leads to different curvature properties encoded in the curvature tensors: Riemann-Christoffel curvature tensor and its contractions. Hence, choosing a gauge has phenomenological consequences. This is independent of the fact that there are the gauge invariant quantities which are the same in any gauge.


The main goal of the present paper is to investigate the possible phenomenological consequences of gauge invariant theories of gravity over WIG space $\tilde W^\text{int}_4$, which are given by the following gravitational Lagrangian:

\bea {\cal L}_\text{wig}=\sqrt{-g}\frac{\phi^2}{12}R=\sqrt{-g}\left[\frac{\phi^2}{12}\hat R+\frac{1}{2}(\der\phi)^2\right],\label{wig-lag}\eea where in the definitions of the relevant geometrical quantities in equations \eqref{gen-aff-c}-\eqref{curv-decomp} one has to make the identification $Q_\mu=2\der_\mu\ln\phi$. This gauge invariant theory of gravity is coupled to the SMP fields via the modified Higgs Lagrangian,

\bea {\cal L}_H=-\frac{\sqrt{-g}}{2}\left[|D^*_gH|^2+\lambda'\left(|H|^2-\kappa^2\phi^2\right)^2\right],\label{ginv-higgs-lag}\eea where $\kappa$ is a dimensionless constant. As demonstrated in \cite{quiros-arxiv-2022}, this is the only possible gauge invariant (classical) theory of gravity with the correct low-curvature gravitational spectrum and where both null and timelike matter fields can be coupled to gravity. This means that gauge symmetry must be an actual symmetry of our past, present and, perhaps, future Universe. We shall show how gauge freedom may explain several outstanding puzzles in the forefront of current science. As we shall see gauge freedom may offer a completely new alternative explanation of several fundamental problems in cosmology, such as: i) the cosmological constant problem (CCP) \cite{ccp-weinberg, ccp-peebles, ccp-padman, ccp-zlatev, ccp-carroll}, ii) the accelerated pace of the cosmic expansion \cite{ccp-peebles, riess-1998, perlmutter-1999, riess-2004, copeland-rev, nojiri-2007, riess-2007, frieman-2008, bamba-2012, suzuki-2012}, iii) the Hubble constant tension \cite{divalentino-rev} and other puzzles. 


This paper has been organized in the following way. In section \ref{sect-wig} we expose the basical aspects of the gauge invariant theory of gravity we shall explore here. In section \ref{sect-geod} we discuss on the motion of point-like test particles in WIG space. These follow autoparallels (same as the geodesics) of WIG space. Here we show why null geodesics respond only to the Riemann curvature of spacetime. The gauge freedom property in connection with gauge symmetry is discussed in section \ref{sect-gauges}. Two outstanding solutions of the present theory are derived in section \ref{sect-sols}. One of these solutions: the de Sitter space, allows us to show that the CCP does not arise in this approach. The phenomenological aspects of the present gauge invariant theory over WIG space are investigated in sections \ref{sect-z}-\ref{sect-flatness}. In section \ref{sect-z} we explain how not only the spacetime curvature but also the nonmetricity affect the gravitational redshift of frequencies. Then, in section \ref{sect-z-q}, it is demonstrated how the accelerated expansion/DE issue is avoided in the present approach, thanks to the impact of nonmetricity on the cosmological redshift. One of the most controversial aspects of gauge symmetry: gauge freedom and gauge fixing, is approached in section \ref{sect-many-w} from the point of view of the many-worlds interpretation of quantum physics. In section \ref{sect-observ} it is demonstrated how the observational evidence may pick out one given gauge (the one where we and the remaining fields of the SMP live) among the infinity of possible gauges. In section \ref{sect-hubble-t} we demonstrate that the Hubble tension issue is a consequence of an incorrect choice of gauge, while in section \ref{sect-flatness} it is shown that the flatness, horizon and relict particles abundance problems do not arise in the present gauge invariant framework. The results of this paper and their possible impact are discussed and brief conclusions are given in section \ref{sect-discu}.
 

Unless otherwise stated, here we use natural units where $\hbar=c=1$ and the following signature of the metric is chosen: $(-+++)$. Greek indices run over spacetime $\alpha,\beta,...,\mu,...=0,1,2,3$, while latin indices $i,j,k...=1,2,3$ run over three-dimensional space. Some times the spatial components of a vector $v^i$ will be represented as three-dimensional vectors $\vec{v}$. Bold-type letters ${\bf v}$ will represent four-dimensional (spacetime) vectors instead. Hence, for instance, ${\bf v}=\{v^0,\vec{v}\}$. Also useful is the following notation. For arbitrary quantities $V_\mu$, $U_\mu$ and $R_{\mu\lambda\nu\sigma}$ with tensorial indexes, the symmetrization and anti-symmetrization of two given indexes are defined as,

\bea &&V_{(\mu}U_{\nu)}:=\frac{1}{2}\left(V_\mu U_\nu+V_\nu U_\mu\right),\nonumber\\
&&R_{(\mu|\lambda|\nu)\sigma}:=\frac{1}{2}\left(R_{\mu\lambda\nu\sigma}+R_{\nu\lambda\mu\sigma}\right),\label{symm}\eea and

\bea &&V_{[\mu}U_{\nu]}:=\frac{1}{2}\left(V_\mu U_\nu-V_\nu U_\mu\right),\nonumber\\
&&R_{\mu\nu[\lambda\sigma]}:=\frac{1}{2}\left(R_{\mu\nu\lambda\sigma}-R_{\mu\nu\sigma\lambda}\right),\label{anti-symm}\eea respectively. Standard Riemann space, which is characterized by vanishing nonmetricity: $\hat\nabla_\alpha g_{\mu\nu}=0$, is denoted by $V_4$.



\section{Gauge invariant theory of gravity in WIG space}\label{sect-wig}


In \cite{quiros-arxiv-2022} it was shown that radiation and massless fields -- the only matter degrees of freedom that couple to gravity in the gauge invariant framework given by \eqref{g-lag} -- do not interact with vectorial nonmetricity. Geometrically this means that the matter fields follow null-geodesics of Riemann space $V_4$, hence, the Weyl gauge vector $Q_\mu$ can not take account neither of the DM nor of the DE. Yet, we can not simply dispense with the nonmetricity vector since it affects the background geometry so that it may influence, for instance, the way the universe inflates. An interesting particular case is the one corresponding to the singular value of the coupling parameter $\omega=6$ in \eqref{g-lag}. In this case the vectorial nonmetricity is just an additional freely propagating radiation field in the background Riemann space $V_4$ so that we may safely ignore the nonmetricity vector without affecting the physical consequences of the resulting gauge invariant theory. 

The problem with the theory behind \eqref{g-lag} is that gauge symmetry must be broken down before, or at least simultaneously, with $SU(2)\times U(1)$ symmetry. I. e., Weyl gauge symmetry does not survives after EW symmetry breaking. However, in Weyl integrable space $\tilde W_4^\text{int}$, since the nonmetricity vector amounts to a gradient of a scalar field, we have an opportunity to improve the above issue. This can be done by lifting the gauge scalar field $\phi$ to the category of a geometric field. In other words, we assume that the nonmetricity of WIG space $\tilde W_4^\text{int}$ is given by, 

\bea \nabla_\alpha g_{\mu\nu}=-2\frac{\der_\alpha\phi}{\phi}g_{\mu\nu},\label{wig-nm}\eea i. e. the nonmetricity vector in Eq. \eqref{vect-nm} is a gradient $Q_\mu=2\der_\mu\phi/\phi$. Under this assumption we have that $\phi^2R=\phi^2\hat R-6\phi\hat\nabla^2\phi$ or, equivalently:

\bea \phi^2 R=\phi^2\hat R+6(\der\phi)^2-6\hat\nabla^\mu\left(\phi\der_\mu\phi\right),\label{wig-decomp}\eea where the last term in the RHS, within an action integral amounts to a boundary term that can be omitted. The action of gauge invariant gravity in $\tilde W_4^\text{int}$ space reads

\bea &&S^\text{wig}_g=\frac{1}{2}\int d^4x\sqrt{-g}\frac{\phi^2}{6}R\nonumber\\
&&\;\;\;\;\;\;\;\;=\frac{1}{2}\int d^4x\sqrt{-g}\left[\frac{\phi^2}{6}\hat R+(\der\phi)^2\right].\label{wig-action}\eea 

The most interesting property of the above action is that matter fields, whether massless or with the mass, can be coupled to gravity without breaking the gauge symmetry. Consider the gauge invariant action over WIG space: $S^\text{wig}_\text{tot}=\int d^4x{\cal L}_\text{tot}$, where the overall Lagrangian is given by:

\bea {\cal L}_\text{tot}=\sqrt{-g}\left[\frac{\phi^2}{12}\hat R+\frac{1}{2}(\der\phi)^2+{\cal L}_\chi\right],\label{wig-tot-lag}\eea where ${\cal L}_\chi$ is the Lagrangian of the matter fields collectively denoted by $\chi$. Gravitational coupling of arbitrary matter fields is possible thanks to the property that in WIG space variation of the metric is not independent of variation of the geometric scalar field $\phi$. Actually, due to gradient nonmetricity law \eqref{wig-nm} one has that (see, for instance, Eq. (3) of Ref. \cite{jackiw-2015}):

\bea \delta g_{\mu\nu}=-2\frac{\delta\phi}{\phi}g_{\mu\nu},\;\delta g^{\mu\nu}=2\frac{\delta\phi}{\phi}g^{\mu\nu}.\label{wig-var}\eea This means that variation of the overall Lagrangian \eqref{wig-tot-lag}, yields

\bea \delta{\cal L}_\text{tot}=\frac{\der{\cal L}_\text{tot}}{\der g^{\mu\nu}}\delta g^{\mu\nu}=2\frac{\der{\cal L}_\text{tot}}{\der g^{\mu\nu}}g^{\mu\nu}\frac{\delta\phi}{\phi},\nonumber\eea or

\bea &&\delta_{\bf g}{\cal L}_\text{tot}=\frac{\der{\cal L}_\text{tot}}{\der g^{\mu\nu}}\delta g^{\mu\nu},\nonumber\\
&&\delta_{\phi^2}{\cal L}_\text{tot}=\frac{\der{\cal L}_\text{tot}}{\der g^{\mu\nu}}g^{\mu\nu}\frac{\delta\phi^2}{\phi^2}.\label{var-wig-tot-lag}\eea Hence, since

\bea &&\frac{\der{\cal L}_\text{tot}}{\der g^{\mu\nu}}=\frac{\sqrt{-g}}{2}\left[\frac{\phi^2}{6} G_{\mu\nu}-T^{(\chi)}_{\mu\nu}\right]\nonumber\\
&&\;\;\;\;\;\;\;\;\;\;\;=\frac{\sqrt{-g}}{2}\left\{\frac{\phi^2}{6}\hat G_{\mu\nu}+\der_\mu\phi\der_\nu\phi-\frac{1}{2}g_{\mu\nu}(\der\phi)^2\right.\nonumber\\
&&\;\;\;\;\;\;\;\;\;\;\;\;\;\;\;\;\;\;\;\;\;\;\;\;\left.-\frac{1}{6}\left(\hat\nabla_\mu\hat\nabla_\nu-g_{\mu\nu}\hat\nabla^2\right)\phi^2-T^{(\chi)}_{\mu\nu}\right\},\nonumber\eea variation of the Lagrangian \eqref{wig-tot-lag} with respect to the metric yields the Einstein's EOM,

\bea &&G_{\mu\nu}=\frac{6}{\phi^2}T^{(\chi)}_{\mu\nu}\;\Leftrightarrow\nonumber\\
&&\hat G_{\mu\nu}-\frac{1}{\phi^2}\left(\hat\nabla_\mu\hat\nabla_\nu-g_{\mu\nu}\hat\nabla^2\right)\phi^2\nonumber\\
&&\;\;\;\;\;\;\;+\frac{6}{\phi^2}\left[\der_\mu\phi\der_\nu\phi-\frac{1}{2}g_{\mu\nu}(\der\phi)^2\right]=\frac{6}{\phi^2}T^{(\chi)}_{\mu\nu},\label{wig-einst-eom}\eea where we have taken into account that the Einstein's tensor of $\tilde W^\text{int}_4$ space can be written in terms of LC (Riemannian) quantities according to 

\bea &&G_{\mu\nu}=\hat G_{\mu\nu}-\frac{1}{\phi^2}\left(\hat\nabla_\mu\hat\nabla_\nu-g_{\mu\nu}\hat\nabla^2\right)\phi^2\nonumber\\
&&\;\;\;\;\;\;\;\;\;\;\;\;\;\;\;\;\;\;+\frac{6}{\phi^2}\left[\der_\mu\phi\der_\nu\phi-\frac{1}{2}g_{\mu\nu}(\der\phi)^2\right].\nonumber\eea Meanwhile, variation of \eqref{wig-tot-lag} with respect to the geometric scalar field $\phi$ leads to:

\bea &&\delta_{\phi^2}{\cal L}_\text{tot}=\frac{\sqrt{-g}}{2}\left[\frac{\phi^2}{6}G_{\mu\nu}-T^{(\chi)}_{\mu\nu}\right]g^{\mu\nu}\frac{\delta\phi^2}{\phi^2}\nonumber\\
&&\;\;\;\;\;\;\;\;\;\;\;\;\;=-\frac{\sqrt{-g}}{12}\left[R+\frac{6}{\phi^2}T^{(\chi)}\right]\delta\phi^2,\nonumber\eea or, if write the WIG curvature scalar $R$ in terms of Riemannian/LC quantities:

\bea -\hat R-6\frac{(\der\phi)^2}{\phi^2}+3\frac{\hat\nabla^2\phi^2}{\phi^2}=\frac{6}{\phi^2}T^{(\chi)},\label{wig-kg-eom}\eea which coincides with the trace of the Einstein's EOM \eqref{wig-einst-eom} without requiring vanishing SET trace. In consequence, the geometric gauge scalar $\phi$ is not a dynamical field: it can be chosen at will. Different choices lead to different gauges. This property of the present theory will be discussed in section \ref{sect-gauges}.


\subsection{Continuity equation}


In Ref. \cite{quiros-arxiv-2022} we have demonstrated, also, that the standard continuity equation in background Riemann space $V_4$: $\hat\nabla^\mu T^{(m)}_{\mu\nu}=0$, takes place as well in a gauge invariant theory over Weyl space (this entails nonvanishing vectorial nonmetricity $Q_\mu$,) which is given by the gravitational Lagrangian \eqref{g-lag}. This means that radiation and massless SMP fields -- the only matter fields which couple to gravity in this framework -- follow null geodesics of Reimann space. In the present case we consider a gauge invariant theory over WIG space -- given by the gravitational action \eqref{wig-action} -- which is distinguished by gradient nonmetricity. The continuity equation can be derived in the following way. Let us take the LC covariant divergence of the quantity $\phi^2G_{\mu\nu}$. According to the EOM \eqref{wig-einst-eom} we get that,

\bea &&\hat\nabla^\mu\left(\phi^2G_{\mu\nu}\right)=\hat\nabla^\mu\phi^2\hat G_{\mu\nu}-(\hat\nabla^2\hat\nabla_\mu-\hat\nabla_\mu\hat\nabla^2)\phi^2\nonumber\\
&&\;\;\;\;\;\;\;\;\;\;\;\;\;\;\;\;\;\;\;\;+\frac{\hat\nabla_\nu\phi^2}{2}\left[3\frac{\hat\nabla^2\phi^2}{\phi^2}-6\frac{(\der\phi)^2}{\phi^2}\right],\nonumber\eea or, if consider the equation 

\bea (\hat\nabla^2\hat\nabla_\mu-\hat\nabla_\mu\hat\nabla^2)\phi^2=\hat R_{\mu\nu}\hat\nabla^\nu\phi^2,\nonumber\eea we obtain

\bea &&\hat\nabla^\mu\left(\phi^2G_{\mu\nu}\right)=\frac{\hat\nabla_\nu\phi^2}{2}\left[-\hat R-6\frac{(\der\phi)^2}{\phi^2}+3\frac{\hat\nabla^2\phi^2}{\phi^2}\right].\nonumber\eea Hence, if in this equation substitute \eqref{wig-kg-eom} and take into account the EOM \eqref{wig-einst-eom}: $\phi^2G_{\mu\nu}=6T^{(\chi)}_{\mu\nu}$, we finally obtain the following continuity equation:

\bea \hat\nabla^\mu T^{(\chi)}_{\mu\nu}=\frac{\der_\nu\phi^2}{2\phi^2}T^{(\chi)}=\frac{\der_\nu\phi}{\phi}T^{(\chi)},\label{wig-cont}\eea which can also be written in the equivalent form,\footnote{Since the conformal weight of the matter SET $w=-2$, then its gauge covariant derivative is given by: $$\nabla^*_\alpha T^{(\chi)}_{\mu\nu}=\nabla_\alpha T^{(\chi)}_{\mu\nu}-2\frac{\der_\alpha\phi}{\phi} T^{(\chi)}_{\mu\nu}.$$}

\bea \nabla^\mu_* T^{(\chi)}_{\mu\nu}=0.\label{cons-theor}\eea 

What equation \eqref{wig-cont} is telling us is that radiation fields with $T^{(\chi)}=0$ follow null geodesics of Riemann space $V_4$, meanwhile, SMP fields with nonvanishing $T^{(\chi)}\neq 0$, follow timelike geodesics of WIG space $\tilde W^\text{int}_4$ instead (see section \ref{sect-geod}). These can be obtained from the continuity equation \eqref{cons-theor} written in the following equivalent form:

\bea \nabla^\mu T^{(\chi)}_{\mu\nu}=2\frac{\der^\mu\phi}{\phi}T^{(\chi)}_{\mu\nu}.\label{cons-theor'}\eea Equations \eqref{wig-cont}, \eqref{cons-theor} and \eqref{cons-theor'} represent the same matter conservation equation in WIG space, which has been written in three different but fully equivalent ways.

Given that in the theory depicted by the overall Lagrangian \eqref{wig-tot-lag} over WIG space $\tilde W^\text{int}_4$, all of SMP fields couple to gravity, no matter whether massless or with the mass, gauge symmetry in this theoretical framework may have impact in the past, present and future of the cosmic dynamics.



\subsection{Final remarks}\label{subsect-eom}


The present theory, which is given by the overall Lagrangian \eqref{wig-tot-lag}, is free of the so called second clock effect (SCE) which plagues gravitational theories in Weyl space $\tilde W_4$ \cite{quiros-arxiv-2022, quiros-2022}. It is not classically forbidden from start since matter fields, both massless and with mass, couple to gravity: massless fields couple exclusively to the LC curvature while fields with the mass couple both to the curvature and to the gradient $\der_\mu\phi/\phi$ (i. e. to the nonmetricity). The derived Einstein's equations \eqref{wig-einst-eom} can be rewritten in the familiar form:

\bea G_{\mu\nu}=\frac{1}{m^2_\text{pl}}T^{(\chi)}_{\mu\nu},\label{eom-theor}\eea where we have introduced the reduced (point-dependent) square Planck mass (compare with Eq. \eqref{mass-phi-rel}),

\bea m^2_\text{pl}=\frac{\phi^2}{6}=m^2_\text{pl,0}\left(\frac{\phi}{\phi_0}\right)^2,\label{planck-mass}\eea where $m^2_\text{pl,0}=m^2_\text{pl}(0)=\phi_0^2/6$ and $\phi_0=\phi(0)$ are the values of the Planck mass and of the geometric scalar field at the origin. The EOM \eqref{eom-theor} is manifestly gauge invariant since both, the Einstein's tensor $G_{\mu\nu}$ and the gauge invariant quantity $T^{(\chi)}_{\mu\nu}/m^2_\text{pl}$ have vanishing conformal weight. Hence, neither is transformed by the gauge transformations \eqref{gauge-t} which, in the present case, amount to:

\bea &&g_{\mu\nu}\rightarrow\Omega^2g_{\mu\nu},\;\phi\rightarrow\Omega^{-1}\phi.\label{gauge-t-theor}\eea Here we are assuming that the constants, no matter whether dimensionless or dimensionful or whether fundamental or not, are not changed by the gauge transformations \eqref{gauge-t-theor}. 

In addition to the Einstein's like equation \eqref{eom-theor}, the conservation equation \eqref{cons-theor} takes place (see also the equivalent equations \eqref{wig-cont} and \eqref{cons-theor'}). These are the totality of equations of motion in the present framework. Recall that there is not an independent EOM for the geometric scalar $\phi$.


\section{Motion of test particles in $\tilde W^\text{int}_4$}\label{sect-geod}


In general autoparallels, which are the ``straightest curves'' of the geometry, do not coincide with the geodesics, which are the ``shortest curves'' \cite{poplawski-arxiv, adak-arxiv, obukhov}. There goes a discussion on whether autoparallels or geodesics describe the motion of test particles in $W_4$ spaces with generalized nonmetricity $Q_{\alpha\mu\nu}$ \cite{quiros-2022, adak-arxiv, obukhov}. However, in Weyl space $\tilde W_4$, as well as in its subspace $\tilde W^\text{int}_4$, autoparallels and geodesics coincide, as in GR. Geodesics and autoparalles can be associated exclusively with the motion of spinless point particles. Spinor fields like the fermions obey the Dirac equation in curved background, while extended spinning test bodies obey the Mathisson-Papapetrou-Dixon equations \cite{mathisson, papapetrou, dixon, wald}.


\subsection{Timelike test particles}

In WIG space the ``timelike'' autoparallels are those curves along which the gauge covariant derivative of the tangent four-velocity vector ${\bf u}$, vanishes. Here $u^\mu=dx^\mu/d\tau$ are the coordinate components of ${\bf u}$ and, as long as this does not cause loss of generality, we chose the proper time $\tau$ to be the affine parameter along the autoparallel curve. The conformal weight of the four-velocity vector $w({\bf u})=-1$. In other words, the autoparallel curves satisfy (for a detailed exposition see appendix B of \cite{quiros-arxiv-2022}):

\bea \frac{d^2x^\alpha}{ds^2}+\{^\alpha_{\mu\nu}\}\frac{dx^\mu}{ds}\frac{dx^\nu}{ds}-\frac{\der_\mu \phi}{\phi}h^{\mu\alpha}=0,\label{time-auto-p}\eea where

\bea h^{\mu\alpha}:=g^{\mu\alpha}+u^\mu u^\alpha=g^{\mu\alpha}-\frac{dx^\mu}{ds}\frac{dx^\alpha}{ds},\label{orto-proj}\eea is the orthogonal projection tensor, which projects any vector or tensor onto the hypersurface orthogonal to the four-velocity vector $u^\mu=dx^\mu/d\tau$.

In $\tilde W^\text{int}_4$ space, since the mass is a point-dependent quantity, then $m$ can not be taken out of the action integral. The action integral in $\tilde W_4$ reads: $S=\int mds.$ From this action the following EOM/geodesic equations can be derived:

\bea \frac{d^2x^\alpha}{ds^2}+\{^\alpha_{\mu\nu}\}\frac{dx^\mu}{ds}\frac{dx^\nu}{ds}-\frac{\der_\mu m}{m}h^{\mu\alpha}=0,\label{time-geod}\eea where the non-Riemannian term $\propto\der_\mu m/m$ accounts for the variation of mass during parallel transport. Hence, if take into account that\footnote{Let us consider the four-momentum vector ${\bf p}=m{\bf u}$, where $m$ is the mass of the point particle. Since under \eqref{gauge-t} the point mass transforms like $m\rightarrow\Omega^{-1}m$, i. e. it has a conformal weight $w(m)=-1$, the weight of the four-momentum $w({\bf p})=-2$. Hence, if apply the law of parallel transport to the four-momentum $$\frac{D^*{\bf p}}{d\tau}=u^\mu\nabla^*_\mu{\bf p}=u^\mu\left(\nabla^*_\mu m\right){\bf u}+mu^\mu\nabla^*_\mu{\bf u}=0,$$ it follows that, $$u^\mu\nabla^*_\mu m=0\;\Rightarrow\;u^\mu\left(\nabla_\mu m-m\frac{\der_\mu\phi}{\phi}\right)=0,$$ or, if take into account that $Dm/d\tau=dm/d\tau=u^\mu\nabla_\mu m$, we obtain that $$\frac{dm}{d\tau}=mu^\mu\frac{\der_\mu\phi}{\phi}\;\Rightarrow\;\frac{dm}{m}=\der_\mu\ln\phi dx^\mu.$$}

\bea \frac{dm}{m}=\der_\mu\ln\phi dx^\mu\;\Rightarrow\;\frac{\der_\mu m}{m}=\frac{\der_\mu\phi}{\phi}.\label{mass-autop}\eea This shows that timelike autoparallels \eqref{time-auto-p} and timelike geodesics \eqref{time-geod} coincide in $\tilde W^\text{int}_4$ space. Eq. \eqref{mass-autop} can be readily integrated to get

\bea m({\bf x})=\mu_0\phi({\bf x})=m_0\frac{\phi}{\phi_0},\label{mass-phi-rel}\eea where $\mu_0$ is a dimensionless integration constant, while $\phi_0=\phi(0)$ and $m_0=m(0)$ are the magnitudes of the mass and of the geometric scalar field evaluated at the origin, respectively.


\subsection{Null particles and fields}

In $\tilde W^\text{int}_4$ the ``null'' autoparallels are those curves along which the gauge covariant derivative of the wave vector ${\bf k}$ with components $k^\mu:=dx^\mu/d\lambda$, vanishes. Here $\lambda$ is a parameter along the null autoparallel and we have to take into account that the conformal weight of the wave vector $w({\bf k})=-2$. Hence, the autoparallel null curves satisfy:

\bea \frac{dk^\alpha}{d\lambda}+\{^\alpha_{\mu\nu}\}k^\mu k^\nu=0.\label{null-auto-p}\eea In other words: photons and radiation in general do not interact with the nonmetricity, i. e., with the gauge vector $Q_\alpha$ (in the present case $Q_\alpha=2\der_\alpha\phi/\phi$.) 

The null geodesic equations can be derived from the following action:

\bea S_\text{null}=\frac{1}{2}\int g_{\mu\nu}\dot x^\mu\dot x^\nu d\xi,\label{action-null-geod}\eea where the dot accounts for derivative with respect to the parameter $\lambda$ of the path $x^\mu(\lambda)$ followed by null fields. From \eqref{action-null-geod} the GR null geodesic equations are obtained. These coincide with the null autoparallels Eq. \eqref{null-auto-p}. Hence, the null geodesic equations do not depend on nonmetricity. This means that, as demonstrated long ago (see appendix D of \cite{wald-book},) photons and radiation interact only with the metric field, i. e., with the LC curvature of spacetime. These do not interact with the gauge scalar $\phi$ whose gradient is the nonmetricity vector.


\section{Gauge freedom}\label{sect-gauges}


In general there is not an independent equation for the geometric scalar $\phi$. This is a direct consequence of gauge freedom since, in addition to the four degrees of freedom to make spacetime diffeomorphisms, we have an additional degree of freedom to make gauge transformations \eqref{gauge-t-theor}. Different choices of the function $\phi({\bf x})$ lead to different gauges of the theory \eqref{eom-theor}, \eqref{cons-theor}. Otherwise, one may fix one of the independent components of the metric, leaving $\phi$ as an independent degree of freedom. In this case one have to solve a differential equation on $\phi$ that is obtained by taking the trace of Einstein's equation Eq. \eqref{eom-theor}: $-R=T^{(\chi,*)}/m^2_\text{pl,0}$ or, if take into account the Riemannian decomposition \eqref{wig-kg-eom}:

\bea \hat\nabla^2\phi^2-2(\der\phi)^2-\frac{\phi^2}{3}\hat R=\frac{\phi^2}{3m^2_\text{pl,0}}T^{(\chi,*)}.\label{kg-theor}\eea Nevertheless, it is recommended to gauge the geometric field $\phi$ away since, otherwise, it would be a ghost degree of freedom due to the incorrect sign of the kinetic energy density term. Hence, the propagating gravitational interactions are the two polarizations of the graviton, as in GR. As a consequence the gravitational coupling coincides with the measured Newton's constant $G_N=3/4\pi\phi^2$.

Different choices of either the gauge scalar $\phi$ or one of the metric functions $g_{\mu\nu}$, with the remaining degrees of freedom completely determined by the equations of motion, lead to physically equivalent gauges in the sense that the same laws of gravity \eqref{eom-theor}, \eqref{cons-theor} are satisfied in any gauge. Nevertheless, each gauge represents a different gravitational theory. These theories are related by the gauge transformations \eqref{gauge-t-theor}. In consequence what we have is a class of conformal equivalence of theories. Gauge fixing the geometric scalar field amounts to choosing a specific theory in the equivalence class. A clear indication that a specific choice $\phi$ picks out a theory of gravity, which is distinguished from any other theory in the conformal equivalence class, is the fact that the measured gravitational constant $G_N=3\phi^{-2}/4\pi$ is directly linked to the scalar field $\phi$, so that the gauge choice is subject to experimental check.


\subsection{General relativity gauge}\label{subsect-gr-gauge}

One of the most outstanding gauges in the present theory is the GR one. It is defined by the following choice of the gauge scalar:

\bea \phi=\phi_0=\text{const}.\label{gr-gauge}\eea 

Given that in the present case the nonmetricity vanishes $Q_\mu=2\der_\mu\phi_0/\phi_0=0$, the WIG space $\tilde W^\text{int}_4$ is replaced by Riemann space $V_4$: $\tilde W^\text{int}_4\rightarrow V_4$. The gauge is characterized by constant mass of timelike point particles: $m=m_0$. Under the choice \eqref{gr-gauge} the gravitational EOM \eqref{eom-theor} reads

\bea \hat G_{\mu\nu}=\frac{1}{m^2_\text{pl,0}} T^{(\chi)}_{\mu\nu},\label{eom-v4}\eea while the continuity equation \eqref{cons-theor} transforms into the GR continuity equation,

\bea \hat\nabla^\lambda T^{(\chi)}_{\lambda\mu}=0,\label{cons-v4}\eea where, as before, a hat over a quantity means that it is defined with respect to the LC connection \eqref{lc-aff-c}. The resulting theory is Einstein's GR in $V_4$. 

Although the manifest symmetry of our gauge invariant theory, that is represented by equations \eqref{eom-theor} and \eqref{cons-theor}, is lost once the GR gauge is fixed, gauge symmetry is implicit in the specific transformations \eqref{gauge-t-theor} that link any gauge with every other one. In this regard the GR gauge is not an exception: it can be linked with any other -- in principle arbitrary -- gauge through the Weyl rescalings \eqref{gauge-t-theor}. Let $\left({\cal M}_4,g_{\mu\nu},\phi\right)\in\tilde W^\text{int}_4$ be a WIG space. Under the gauge transformation \eqref{gauge-t-theor},

\bea g_{\mu\nu}\rightarrow\Omega^2 g_{\mu\nu},\;\phi\rightarrow\Omega^{-1}\phi_0,\nonumber\eea or, equivalently, under

\bea g_{\mu\nu}\rightarrow\left(\frac{\phi_0}{\phi}\right)^2g_{\mu\nu},\label{gauge-t-0}\eea one can map the gravitational theory -- defined by the EOM \eqref{eom-theor} and \eqref{cons-theor} --  over $\tilde W^\text{int}_4$ space, into GR which is driven by the EOM \eqref{eom-v4} and \eqref{cons-v4}, over Riemannian manifold $V_4$:

\bea \left[\left({\cal M}_4,g_{\mu\nu},\phi\right)\in\tilde W^\text{int}_4\right]\rightarrow\left[\left({\cal M}_4,g_{\mu\nu},\phi_0\right)\in V_4\right].\label{map}\eea The converse is also true: under the inverse gauge transformation

\bea g_{\mu\nu}\rightarrow\left(\frac{\phi}{\phi_0}\right)^2g_{\mu\nu},\label{gauge-t-0-inv}\eea one can map general relativity over Riemann space $V_4$, into manifest gauge invariant gravitational theory over WIG space $\tilde W^\text{int}_4$,

\bea \left[\left({\cal M}_4,g_{\mu\nu},\phi_0\right)\in V_4\right]\rightarrow\left[\left({\cal M}_4,g_{\mu\nu},\phi\right)\in\tilde W^\text{int}_4\right].\label{inv-map}\nonumber\eea Hence, GR belongs in the class of conformal equivalence of our gauge invariant theory \eqref{wig-tot-lag}. In other words, GR theory is just one of the feasible gauges in the infinite equivalence class: although GR is not a manifest gauge invariant theory, it belongs in a bigger gauge invariant theory. Even if the different gauges are physically equivalent in the sense that the same gravitational laws \eqref{eom-theor}, \eqref{cons-theor} are satisfied, each gauge yields a different geometrical (and physical) description of the world. 

We want to underline that different constant values $\phi_{i0}$ ($i=1,2,...,N$ with $N\rightarrow\infty$) lead to different copies of GR theory. Each copy has a different value of the measured square Planck mass $M^2_\text{pl,i}=\phi^2_{i0}/6$, and also different value of the mass parameter $v_{i0}=\kappa\phi_{i0}$ in the SMP Lagrangian \eqref{ginv-higgs-lag}, so that each particle of the SMP acquires a different mass in the different copies: $m_i=\mu_{i0}\phi_{i0}$, where $\mu_{i0}=\kappa g_{iY}$ is a dimensionless constant ($g_{iY}$ is the Yukawa coupling.) Hence, the GR gauge is a in principle infinite set of $i$ copies of general relativity with different values of several universal constants $\{M^2_\text{pl,i},\;v_{i0},...\}$. Other constants such as the Planck constant $\hbar$, the speed of light $c$, the electron charge $e$, etc. are the same in all $i$ copies of GR theory.


\section{Outstanding solutions}\label{sect-sols}


Due to their simplicity and importance here we shall discuss on two relevant solutions of the EOMs \eqref{eom-theor} and \eqref{cons-theor} which are special in some sense.


\subsection{Minkowski space}

In this case we fix the metric functions without invoking the EOM \eqref{eom-theor}, while the gauge scalar can be found with the help of the equation \eqref{kg-theor}. Minkowski space is one of the simplest solutions of any theory of gravity. In the present case its simplicity allows us to qualitatively discuss on certain phenomenological consequences of the theory.

According to this solution the gravitational effects are described in flat Minkowski space with metric $\eta_{\mu\nu}=\text{diag}(-1,1,1,1)$ by the gauge scalar exclusively. The equations of motion \eqref{cons-theor} and \eqref{kg-theor}, simplify to

\bea \der^\lambda T^{(\chi)}_{\lambda\mu}=\frac{\der_\mu\phi}{\phi}T^{(\chi)},\label{mink-cons}\eea and to

\bea \der^2\phi=\frac{T^{(\chi)}}{\phi},\label{mink-kg}\eea respectively, where we adopted the following notation: $\der^2\equiv\eta^{\mu\nu}\der_\mu\der_\nu$. 

While in Riemannian spacetimes the Minkowski metric corresponds to vacuum solution, in $\tilde W^\text{int}_4$ space it is not associated with vacuum as it can be seen from equations \eqref{mink-cons} and \eqref{mink-kg}. These equations are fully equivalent to the equations of Nordstr$\ddot{\text{o}}$m's theory of gravity \cite{deruelle-2011}. This theory is ruled out by Solar system's experiments \cite{will-rev}. In particular, as seen from the null geodesic equation \eqref{null-auto-p} which, in the present case amounts to $dk^\mu/d\lambda=0$, there is no light-bending in this theory. The redshift of frequencies may be explained as due to point-dependent property of the mass expressed in Eq. \eqref{mass-phi-rel}.


\subsection{de Sitter space and the CCP}\label{subsect-ccp}

In the present theoretical framework the de Sitter space is defined by,

\bea R_{\mu\nu}=\frac{R}{4}g_{\mu\nu}=\Lambda_\text{eff}\,g_{\mu\nu}\;\Rightarrow\;G_{\mu\nu}=-\Lambda_\text{eff}\,g_{\mu\nu},\label{dsitter-space}\eea where the effective cosmological constant $\Lambda_\text{eff}$ is dynamical. Actually, since the product $\Lambda_\text{eff}\,g_{\mu\nu}$ must be gauge invariant and, given that under \eqref{gauge-t-theor} $g_{\mu\nu}\rightarrow\Omega^2 g_{\mu\nu}$, then $\Lambda_\text{eff}\rightarrow\Omega^{-2}\Lambda_\text{eff}$. This means that we can write $\Lambda_\text{eff}=\Lambda\phi^2$, where $\Lambda$ is a dimensionless free constant. In consequence, the EOM \eqref{eom-theor} transforms into,

\bea G_{\mu\nu}=\frac{6}{\phi^2}T^\text{vac}_{\mu\nu}=-\Lambda\phi^2g_{\mu\nu},\label{vac-feq}\eea where the SET of vacuum is given by:

\bea T^\text{vac}_{\mu\nu}=-\frac{\Lambda}{6}\phi^4g_{\mu\nu}.\label{vac-set}\eea The gauge invariant continuity equation \eqref{cons-theor} reads,

\bea \nabla^\lambda T^\text{vac}_{\lambda\mu}=2\frac{\nabla^\lambda\phi}{\phi}\,T^\text{vac}_{\lambda\mu}.\label{cont-eq}\eea 

From Eq. \eqref{vac-set} it follows that the energy density of vacuum depends on the spacetime point:

\bea \rho_\text{vac}=\frac{\Lambda}{6}\,\phi^4=\rho_\text{vac}^0\left(\frac{\phi}{\phi_0}\right)^4,\label{vac-edensity}\eea where $\rho_\text{vac}^0=\rho_\text{vac}(0)=\Lambda\phi_0^4/6$ is the energy density of vacuum at the origin. As we shall see this dynamical behavior may explain the CCP \cite{ccp-weinberg, ccp-peebles, ccp-padman, ccp-zlatev, ccp-carroll}. 

Let us consider a FRW spacetime with flat spatial sections which is given by the line element,

\bea ds^2=-dt^2+a^2(t)\delta_{ij}dx^i dx^j,\;\;\;\;i,j=1,2,3,\label{frw-met}\eea where $a(t)$ is the dimensionless scale factor and $t$ is the cosmic time. According to Eq. \eqref{vac-edensity} the energy density of vacuum is a function of the cosmic time $t$,

\bea \rho_\text{vac}(t)=\rho_\text{pl}\left[\frac{\phi(t)}{\phi_\text{pl}}\right]^4,\label{vac-edens}\eea where we have conveniently shifted the coordinate origin to the Planck time $t_\text{pl}$ elapsed after the bigbang at $t=0$, so that:

\bea \rho_\text{pl}=\rho_\text{vac}(t_\text{pl}),\;\phi_\text{pl}=\phi(t_\text{pl}),\nonumber\eea are the Planck energy density and the value of the gauge scalar at Planck time. We assume next that $\phi(t)$ is a monotonically decreasing function:\footnote{Since $\phi$ is a gauge field (we can replace it by any function of the cosmic time) this assumption does not affect the generality of our analysis.} $\phi(t)<\phi_\text{pl}$ for any $t>t_\text{pl}$. At present time $t=t_0$ the energy density of vacuum reads

\bea \rho_\text{vac}(t_0)=\rho_\text{pl}\left[\frac{\phi(t_0)}{\phi_\text{pl}}\right]^4.\nonumber\eea According to the cosmological data the present measured value of the vacuum energy density $\rho_\text{vac}(t_0)$ is about $120$ orders of magnitude smaller than its Planck's value:

\bea \rho_\text{vac}(t_0)\approx 10^{-120}\rho_\text{pl}.\label{ccp-comp}\eea This means that a reasonable difference of about $30$ orders of magnitude between the present value of the geometric scalar and its value at Planck time: $\phi(t_0)\sim 10^{-30}\phi_\text{pl}$, resolves the apparent discrepancy between the measured value of the vacuum energy density and its computed value according to well-motivated estimates \cite{ccp-weinberg}. This is not a unreasonable requirement since, according to conservative estimates, during inflation the scale factor is increased by a factor of at least $10^{27}$.


\section{Redshift of frequency}\label{sect-z}


In this section we shall show that the overall redshift effect in the gravitational theory described by the EOMs \eqref{eom-theor} and \eqref{cons-theor}, which is built on $\tilde W^\text{int}_4$ space, is the result of the combined effect of the cosmological redshift of frequencies -- due to photon's propagation in a background spacetime with nonvanishing curvature -- and of an additional redshift which results from spacetime variation of masses. 

We shall consider the FRW metric \eqref{frw-met}. In the GR gauge the FRW line element reads,

\bea d\hat s^2=-dt^2_\text{GR}+a^2_\text{GR}\delta_{ij}dx^i dx^j,\label{hat-frw-met}\eea where, according to Eq. \eqref{gauge-t-0}, $$dt_\text{GR}=\Omega dt,\;a_\text{GR}=\Omega a,\;\Omega=e^{(\vphi-\vphi_0)/2}.$$ Notice that the cosmic time transforms like the line-element under the gauge transformations \eqref{gauge-t-theor}.


\subsection{Redshift due to photon propagation in a curved spacetime}

According to the null geodesic equation \eqref{null-auto-p}, photons and radiation do not interact with the gauge scalar $\phi$. This means that these are not able to ``see'' the WIG structure of $\tilde W^\text{int}_4$ space. Instead, these see the Riemannian structure of $V_4$ space, just as in general relativity. If write Eq. \eqref{null-auto-p}, in FRW space we have that,

\bea \frac{d\omega}{d\xi}+a\dot{a}\delta_{ij}k^i k^j=0,\label{frw-null-geod}\eea where, for any function $\chi=\chi(t)$, $\dot\chi\equiv d\chi/dt$, $\xi$ is the affine parameter along the photon's path and $k^\mu=dx^\mu/d\xi$ is the four-wave vector, with components: $k^0=dt/d\xi=\omega$ (photon's cyclic frequency) and $k^i=dx^i/d\xi$. The following expression is satisfied by the wave vector:

\bea g_{\mu\nu}k^\mu k^\nu=0\;\Rightarrow\;-\omega^2+a^2(t)\delta_{ij}k^i k^j=0.\label{k2}\eea If substitute this equation back into \eqref{frw-null-geod} we get that,

\bea \frac{d\omega}{d\xi}+\frac{\dot{a}}{a}\omega^2=0,\nonumber\eea or, if recall that $\omega=2\pi\nu$ ($\nu$ is the photon's frequency) and that $\omega=dt/d\xi$ $\Rightarrow\dot{a}\omega=da/d\xi$, then the null geodesic equation can be written in the following way:

\bea \frac{d\nu}{d\xi}+\frac{1}{a}\frac{da}{d\xi}\nu=0.\label{frw-null-geod'}\eea Straightforward integration of this equation leads to the redshift of photon's frequency:

\bea \nu=\frac{\nu_0}{a},\;\nu_0\equiv e^C,\label{gr-redshift}\eea where $C$ is an integration constant. This is the GR redshift effect that is due to the propagation of photons (and radiation in general) in a curved spacetime background. 

Let us assume that a photon with frequency $\nu_\text{em}=\nu(t)$ is emitted at some time $t$ into the past. After propagating in a given FRW background, the photon will have a frequency $\nu_\text{abs}=\nu(0)$ at some present time $t=0$, when it is absorbed. Hence, we may define the (relative) cosmological redshift of frequency, which is originated by the effect of the curvature of spacetime on the photon during its propagation, in the following way:

\bea z_\text{curv}\equiv\frac{\nu_\text{em}-\nu_\text{abs}}{\nu_\text{abs}}=\frac{\nu_\text{em}}{\nu_\text{abs}}-1=\frac{a(0)}{a(t)}-1,\label{z-curv}\eea where $a$ is the scale factor in a FRW-$\tilde W^\text{int}_4$ space with metric \eqref{frw-met}. In the GR gauge $z_\text{curv}=z_\text{GR}$. Actually, this is the same as the standard GR redshift of frequency:

\bea z_\text{GR}=\frac{\nu_\text{em}}{\nu_\text{abs}}-1=\frac{a_\text{GR}(0)}{a_\text{GR}(t_\text{GR})}-1,\label{z-gr}\eea where $a_\text{GR}(0)$ is the scale factor evaluated at present time $t_\text{GR}=0$. Below we shall consider the normalization where $a_\text{GR}(0)=1$.


\subsection{Redshift due to mass variation}

In addition to the redshift effect \eqref{z-curv}, which is originated from propagation of light in curved $\tilde W^\text{int}_4$ spacetime, an additional redshift arises which is basically due to nonmetricity. It is based on point-dependent property of the mass parameter $m=m({\bf x})$, given that photons are emitted and absorbed by atoms. Let us consider, for instance, the hydrogen atom. In the hydrogen atom the energy of each energy level, labeled by $n$, is given by:

\bea E_n=-\frac{m_e\alpha^2}{2n^2},\label{e-level}\eea where $m_e$ is the mass of the electron and $\alpha\approx 1/137$ is the fine-structure constant. Any changes in the mass $m_e$ over spacetime will cause changes in the energy levels and, consequently, in the energy of the atomic transitions between states $n_i$ and $n_f$,

\bea \nu_{if}=|E_{n_f}-E_{n_i}|=\frac{m_e\alpha^2}{2}\left|\frac{1}{n^2_f}-\frac{1}{n^2_i}\right|.\label{atm-t}\eea Hence, the frequency of either emitted or absorbed photons will be affected by the variation of the electron mass over spacetime, which is given by Eq. \eqref{mass-phi-rel}:

\bea m_e({\bf x})=m_{e0}\frac{\phi({\bf x})}{\phi_0},\label{e-mass}\eea where $m_{e0}$ is the measured mass of the electron at the origin.

Let us imagine that an hydrogen atom, which is placed at point ${\bf x}\equiv\{x^\mu\}$, emits a photon with frequency:

\bea \nu_{if}({\bf x})=\frac{m_{e0}\alpha^2}{2}\left|\frac{1}{n^2_f}-\frac{1}{n^2_i}\right|\frac{\phi({\bf x})}{\phi_0},\label{freq-x}\eea while a second hydrogen atom, which is placed at the coordinate origin ${\bf x}=0$, absorbs the emitted photon. In order for the photon to be absorbed by the atom at the origin, its frequency must equal,

\bea \nu_{if}(0)=\frac{m_{e0}\alpha^2}{2}\left|\frac{1}{n^2_f}-\frac{1}{n^2_i}\right|.\label{freq-0}\eea This causes an additional cosmological redshift of frequency which is due to the nonmetricity,

\bea z_\text{nm}=\frac{\nu_{if}({\bf x})}{\nu_{if}(0)}-1=\frac{\phi({\bf x})}{\phi_0}-1,\label{z-add}\eea so that it does not arise in general relativity. Recall that in GR the masses of particles (including atoms, etc.) are constants, so that $\phi({\bf x})=\phi_0$ $\Rightarrow z_\text{nm}=0$.

The overall redshift of frequency in $\tilde W^\text{int}_4$ space equals,

\bea z_\text{tot}=z_\text{curv}+z_\text{nm},\label{z-tot}\eea where $z_\text{curv}$ and $z_\text{nm}$ are given by \eqref{z-curv} and \eqref{z-add}, respectively. In a FRW cosmological scenario Eq. \eqref{z-tot} can be written in the following way:

\bea z_\text{tot}(t)=\frac{\phi(t)}{\phi(0)}\left[\frac{a(0)\phi(0)}{a(t)\phi(t)}+1\right]-2,\label{z-tot-frw}\eea where $t$ is the cosmic time. Notice that the product $a\phi$ is a gauge invariant quantity since, under \eqref{gauge-t-theor}: $a\rightarrow\Omega a$ and $\phi\rightarrow\Omega^{-1}\phi$. This means that the following equality takes place:

\bea a_\text{GR}=a\phi,\label{equal}\eea where $a_\text{GR}$ is the scale factor in the GR gauge in the normalization where $\phi_0=1$.


\section{Redshift and accelerated expansion}\label{sect-z-q}


In this section and in other sections of this paper where simplicity of mathematical handling is required, we make the scalar field replacement $\phi=e^{\vphi/2}$, or, the more appropriate:

\bea \frac{\phi({\bf x})}{\phi_0}=e^{\frac{\vphi({\bf x})-\vphi_0}{2}}.\label{phi-vphi}\eea This means that negative $\phi$-s are not allowed. However, this does not spoil the generality of our analysis. For instance, a normalization where $\vphi(0)=0$ means that $\phi(0)=1$, and so on. We should notice that the geometric field $\phi$ has mass units, while in the equation $\phi=e^{\vphi/2}$ it appears as a dimensionless field. This can be improved by using the relationship \eqref{phi-vphi}, which is free of the mentioned issue.

Let us explain how the accelerated pace of the cosmic expansion can be explained (avoided) in our theory. As we shall see this is the result of the overall redshift of frequency which, in the present case, is the result of the combined effect of curvature redshift $z_\text{curv}$ and of the additional redshift $z_\text{nm}$, due to spacetime variation of masses in connection with nonmetricity. 

According to \eqref{gauge-t-0-inv}, under the gauge transformations \eqref{gauge-t-theor} the GR (Riemannian) FRW metric transforms like 

\bea -dt^2_\text{GR}+a_\text{GR}^2\delta_{ij}dx^idx^j\;\rightarrow\;\Omega^2(-dt^2+a^2\delta_{ij}dx^idx^j),\nonumber\eea where, according to \eqref{gauge-t-0-inv}: $\Omega^2=e^{\vphi-\vphi_0}.$ Let us, for simplicity and definiteness, assume the following normalization: $\vphi_0\equiv\vphi(0)=0$ and $a_\text{GR}(0)=a(0)=1$. The GR scale factor obeys Eq. \eqref{equal}: $a_\text{GR}=e^{\vphi/2}a,$ where $a(t)$ is the time dependent scale factor in any other gauge. But, according to Eq. \eqref{z-gr} we have that,

\bea a_\text{GR}(t)=\left(z_\text{GR}+1\right)^{-1}.\label{z-curv-inv}\eea Hence, it follows that,

\bea a(t)e^{\vphi(t)/2}=\left(z_\text{GR}+1\right)^{-1}.\label{z-cur-rel}\eea Therefore, equation \eqref{z-tot-frw} can be written as,

\bea z_\text{tot}=e^{\vphi/2}z_\text{GR}+2\left(e^{\vphi/2}-1\right).\label{z-tot-fin}\eea This means that if our theory were correct, the redshift $z_\text{GR}$ appearing in astrophysical data sets and in tables, which is computed under the assumption that GR theory over $V_4$ space is the correct theory of gravity, is to be replaced by the overall redshift in Eq. \eqref{z-tot-fin}: $z_\text{GR}\,\rightarrow\,z_\text{tot}.$ 

What does the above result entails for the physical interpretation of present cosmological and astrophysical data? Let us, for illustration, consider one of the first data sets with evidence for accelerated expansion, from one of the now famous collaborations \cite{perlmutter-1999}. Let us assume also that $\vphi$ is a small quantity, so that Eq. \eqref{z-tot-fin} can be written in the following way,

\bea z_\text{tot}\approx\left[1+\frac{\vphi(z_\text{GR})}{2}\right]z_\text{GR}+\vphi(z_\text{GR}),\label{z-tot-approx}\eea where, for definiteness, we choose the simplest possible function $\vphi=\epsilon z_\text{GR}$ ($\epsilon$ is a small parameter). Hence,

\bea z_\text{tot}=(1+\epsilon)z_\text{GR}+\frac{\epsilon}{2}z^2_\text{GR}.\label{z-master}\eea This will be our master equation to determine the overall (observed) redshift in our theory. Our approximation is not bad for redshifts $z_\text{GR}<1$, so that in the present case it may serve our illustrative purpose.


\begin{figure*}[t!]
\includegraphics[width=7.5cm]{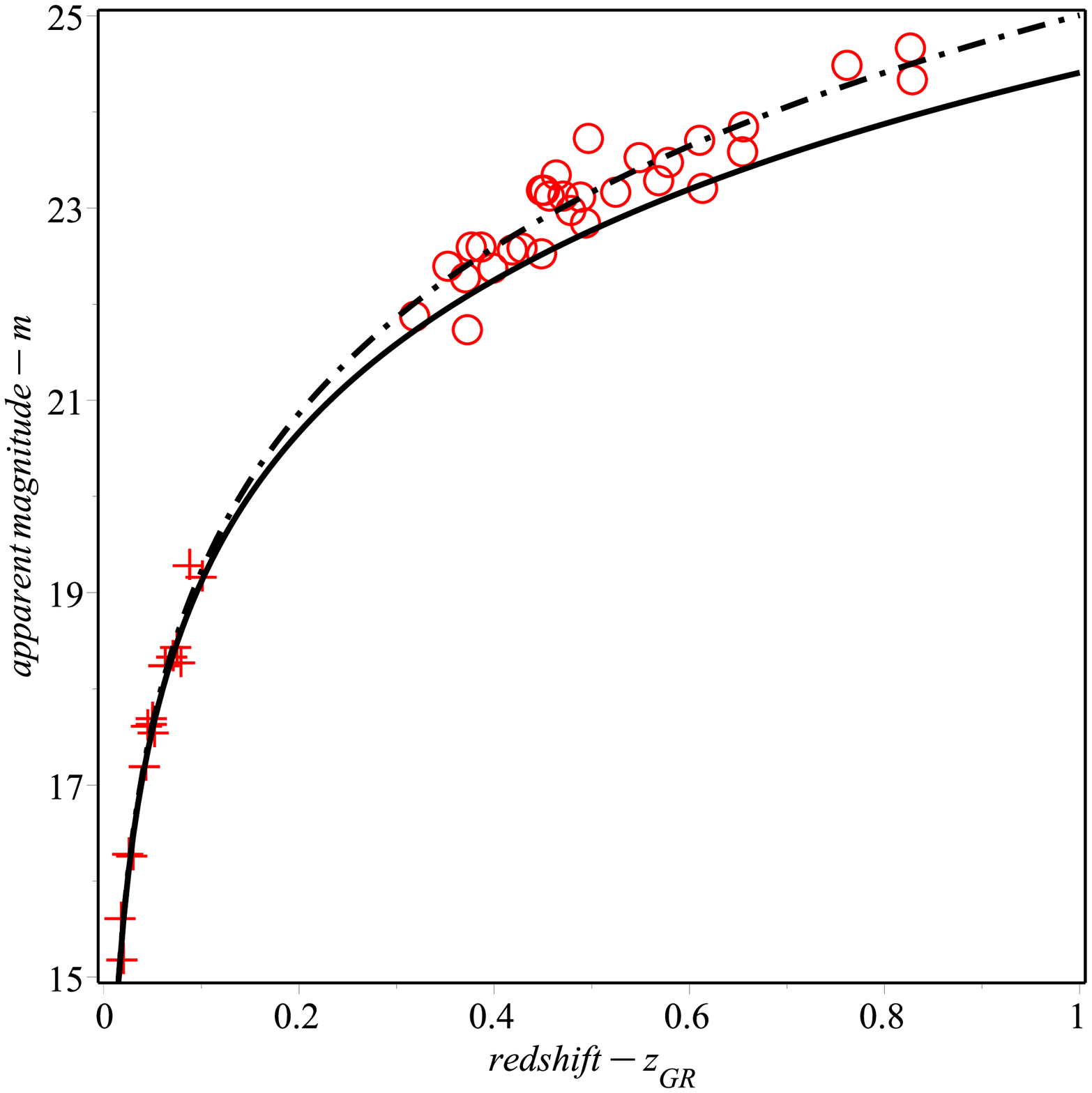}
\includegraphics[width=7.5cm]{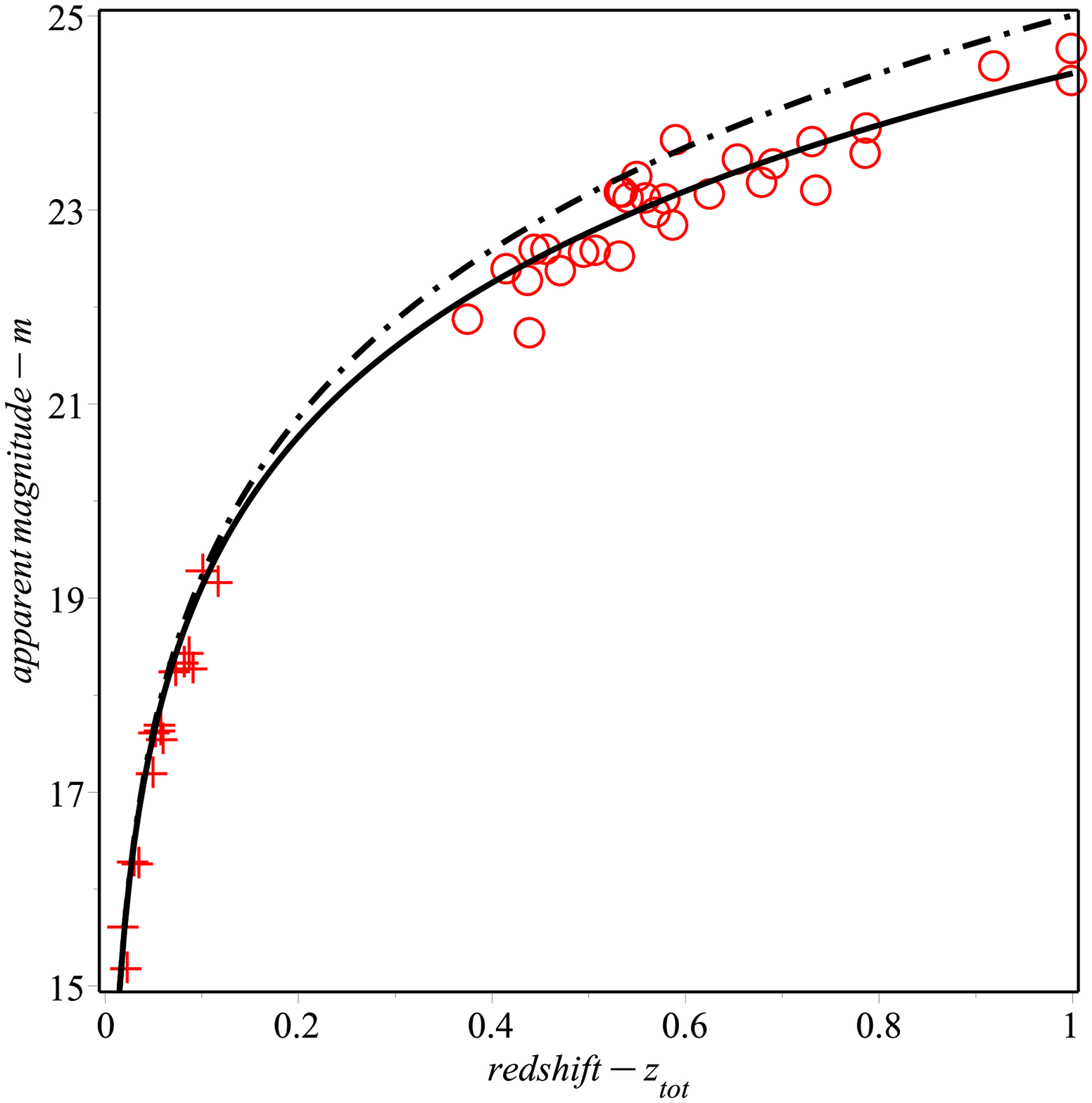}
\vspace{1.3cm}\caption{Plots of the apparent magnitude \eqref{app-m} vs redshift for two choices of the cosmological parameters $\Omega^0_m\equiv\rho_m(0)/3M^2_\text{pl}H^2_0$-present value of the dimensionless energy density of the dark matter and $\Omega^0_\Lambda\equiv\Lambda/3M^2_\text{pl}H^2_0$-present value of the dimensionless energy density of the cosmological constant. We have not included the error bars since the plots are for illustrative purposes. The dash-dot curve corresponds to the choice $\Omega^0_m=0.3$, $\Omega^0_\Lambda=0.7$, while the solid curve is for $\Omega^0_m=1$, $\Omega^0_\Lambda=0$. The crosses and the circles represent observational points corresponding to small-redshift data and to high-redshift data, respectively. Both sets of data are taken from TAB. 1 (high-redshift data) and TAB. 2 (small-redshift data) in Ref. \cite{perlmutter-1999}. We have not included all data points but just a representative number of them. In the left figure we have used the values of the redshift $z_\text{GR}$ computed within GR theory over Riemann space $V_4$, which is the one that appears in TABs. 1 and 2 of Ref. \cite{perlmutter-1999}. In the right figure we have used, instead, the redshift $z_\text{tot}$ given by Eq. \eqref{z-master}, that arises in the present gauge invariant theory under the assumption of small $\vphi(z_\text{GR})=\epsilon z_\text{GR}$ (we have chosen $\epsilon=0.15$). It is seen that, while in standard Riemannian GR the $\Lambda$CDM model is favored by the high-redshift data, in our gauge invariant setup (under the assumed approximation,) the CDM-dominated universe $3H^2=M^{-2}_\text{pl}\rho_m$ with vanishing cosmological constant is favored instead.}\label{fig1}\end{figure*}



\subsection{Accelerating or decelerating expansion?}

Measuring distances to distant stars is not an easy task. Several quantities such as the apparent magnitude $m$, the absolute magnitude $M$ and the luminosity distance $d_L$ are involved. The latter is related with the energy flux $\Phi$ measured by the observer at $z=0$, which comes from a distant source with actual luminosity $L$, in the following way,

\bea d_L=\sqrt\frac{L}{4\pi\Phi}.\nonumber\eea Meanwhile, the absolute and apparent magnitudes $M$ and $m$, are logarithmic measures of luminosity and flux, respectively.

The luminosity distance $d_L$ is, in general, a model dependent quantity. Actually, it can be related to theoretical quantities as it follows:

\bea d_\text{L}=d_\text{C}(1+z),\label{dl}\eea where the comoving distance $d_\text{C}$ is given by

\bea d_\text{C}=\int_t^0\frac{dt'}{a(t')}=\int_0^z\frac{dz'}{H(z')}.\label{co-mov-d}\eea In this equation the expression for the Hubble parameter as a function of the redshift parameter $z$ can be found from the Friedmann equation of the model. For instance, for the $\Lambda$CDM model we have that,\footnote{Although for simplicity here we do not use quantities with the hat, it is implicit that the $\Lambda$CDM model is based on \eqref{eom-v4} which is satisfied only in the GR gauge (see also Eq. \eqref{vac-feq}). In the present case in Eq. \eqref{eom-v4} one have to make the replacement, $T^{(m)}_{\mu\nu}\rightarrow T^{(m)}_{\mu\nu}+T^\text{vac}_{\mu\nu}$.}

\bea H_0d_\text{L}=\frac{1+z}{\sqrt{\Omega^0_m}}\int^z_0\frac{dz'}{\sqrt{(1+z')^3+\Omega^0_\Lambda/\Omega^0_m}},\label{h0dl}\eea where $\Omega^0_m\equiv\Omega_m(0)=\rho_m(0)/3M^2_\text{pl}H^2_0$ is the present value of the dimensionless (normalized) energy density of the cold dark matter (CDM), while $\Omega^0_\Lambda\equiv\Lambda/3M^2_\text{pl}H^2_0$ is the present value of the dimensionless energy density of the cosmological constant $\Lambda$. 

According to Eq. (31) of reference \cite{ccp-padman}, the redshift-dependent apparent magnitude $m$ of distant type IA supernovae is given by\footnote{Since we consider the dimensionless combination $H_0d_L(z)$ rather than $d_L(z)$, this means that in the equation to determine the apparent magnitude a term $5\log_{10}(H_0)$ is to be taken away.}

\bea m(z)=M-5\log_{10}h+42.38+5\log_{10}\left[H_0d_L(z)\right].\nonumber\eea Here we take $M=-19.09$ and $h=0.7$ so that,

\bea m(z)=5\log_{10}(H_0d_L)+24.06,\label{app-m}\eea which is the master equation to determine the theoretical values of the apparent magnitude. 

Let us notice that, since the quantities $d_L$, $M$ and $m$ are related with luminosity and the flux of photons, i. e., with the propagation of photons in spacetime, neither is modified by the nonmetricity (recall that photons do not interact with nonmetricity but only with the curvature of space.) This means that these are gauge invariant quantities (Riemannian null geodesics already satisfy gauge invariant equations) which equal their GR value in any gauge. This is why, in what follows -- without loss of generality -- we compute $m(z)$ in the GR gauge exclusively. What changes from gauge to gauge is the magnitude of the overall redshift factor $z_\text{tot}$, which in the GR gauge amounts to $z_\text{GR}$.

Now we are in position to explain the way in which the accelerated expansion/dark energy issue can be avoided in our theory. As illustration we shall consider one of the first data sets on high redshift supernovae measurements \cite{perlmutter-1999}, which provided early evidence for accelerated expansion of the universe. 

First we shall explore the $\Lambda$CDM model, which arises in the GR gauge of the present gauge invariant theory \eqref{wig-tot-lag}, \eqref{eom-theor}, \eqref{cons-theor}, when we choose $\vphi=0$, with a nonvanishing cosmological constant, for two arrangements of the present values of the dimensionless energy densities $\Omega^0_m$ and $\Omega^0_\Lambda$: $(\Omega^0_m,\Omega^0_\Lambda)=(0.3,0.7)$ and $(\Omega^0_m,\Omega^0_\Lambda)=(1,0)$, respectively. It is well-known that most of the existing data-sets point to a relationship $\Omega^0_m/\Omega^0_\Lambda\approx 0.43$, so that, for instance, the choice where $\Omega^0_m=1$ and $\Omega^0_\Lambda=0$, does not fit well the observational evidence. Second we shall consider the same two choices of the mentioned cosmological parameters but in the gauge where $\vphi(z_\text{GR})=\epsilon z_\text{GR}$ ($\epsilon$ is a small number which, for our illustrative purpose, we choose to be $\epsilon=0.15$.) Since the values of the apparent magnitude, which coincide with their GR values, are the same in all of gauges, the only thing we have to modify in the chosen data set are the values of the redshift parameter which is $z_\text{GR}$ in the GR gauge ($\vphi=0$), but in the alternative gauge where $\vphi=\epsilon z_\text{GR}$, it should be replaced by the overall redshift $z_\text{tot}$ \eqref{z-master}.

The results are shown in FIG. \ref{fig1}, where the dash-dot curve corresponds to theoretical predictions of the apparent magnitude $m(z)$ for the $\Lambda$CDM model with $\Omega^0_m=0.3$ and $\Omega^0_\Lambda=0.7$, while the solid curve corresponds to the choice $\Omega^0_m=1$, $\Omega^0_\Lambda=0$. This last choice amounts to matter-dominated (decelerated) Friedmann expansion: $3H^2=M^{-2}_\text{pl}\rho_m$. A plot of the apparent magnitude $m$ vs curvature redshift $z_\text{GR}$ is shown in the left figure, while in the right figure a plot of the apparent magnitude vs the overall redshift $z_\text{tot}$ is shown. In other words, in the left figure we have a fit of the $\Lambda$CDM model in the GR gauge of our gauge invariant setup, where we set $\vphi=0$ (formally it is just standard general relativity,) to observational data for two different arrangements of the cosmological parameters $\Omega^0_m$ and $\Omega^0_\Lambda$. Meanwhile in the right figure we have a fit of the same model (same arrangements of the cosmological parameters) but in another gauge where $\vphi=\epsilon z_\text{GR}$. While in the GR gauge, where the spacetime background is $V_4$, the redshift $z_\text{GR}$ is due exclusively to propagation of photons in a curved spacetime, in any other gauge where $\vphi=\vphi(t)$ and the background space is $\tilde W^\text{int}_4$, the overall redshift $z_\text{tot}$ gets contributions both from propagation of photons in a curved background and from variation of masses of the atoms from point to point in spacetime.

It is seen in the left figure in FIG. \ref{fig1} that the observational data favors the $\Lambda$CDM model with $\Omega^0_m=0.3$ and $\Omega^0_\Lambda=0.7$. Meanwhile in the right figure the same observational data, interpreted in a different gauge of the gauge invariant theory over $\tilde W^\text{int}_4$ spacetime, namely, $\vphi=\epsilon z_\text{GR}$, seems to favor cosmological evolution dominated by CDM (vanishing cosmological constant). In this last case the supernova data brings no evidence for accelerating expansion, so that the dark energy plays no role in the cosmological dynamics and may be safely ignored. 

The present analysis has been mostly illustrative and it did not involve other data sets than one of the first high-redshift measurements reported in \cite{perlmutter-1999}. Other evidences for accelerating expansion within the framework of Riemannian GR-based $\Lambda$CDM model, such as those related with cosmic microwave background (CMB) temperature anisotropies, baryon acoustic oscillations (BAO), etc., should be carefully analyzed in our present theory before we may come to a definitive (solid) conclusion on the possible abandonment of the dark energy idea.


\section{``Many-worlds'' interpretation of gauge invariance}\label{sect-many-w}


In this section, for convenience, we shall use the geometric scalar field $\phi$ instead of $\vphi$, recalling that these are related by Eq. \eqref{phi-vphi}. In the next section we come back to $\vphi$ again.

The gauge invariant theory explored in previous sections shows a distinctive feature: due to invariance of the EOMs \eqref{eom-theor} and \eqref{cons-theor} under the gauge transformations \eqref{gauge-t-theor}, in addition to the four degrees of freedom to make diffeomorphisms, there is an additional degree of freedom to make gauge transformations. This is reflected in the fact that there is not an independent equation of motion for the gauge scalar $\phi$. Hence, any choice of the function $\phi=\phi({\bf x})$ will satisfy the equations of motion \eqref{eom-theor} and \eqref{cons-theor}. Different choices of the gauge scalar lead to different gauges of the present gauge invariant theory (see the discussion in Sec. \ref{sect-gauges}.) Although the gravitational laws \eqref{eom-theor}, \eqref{cons-theor} are the same in all of possible gauges, each gauge

\bea {\cal G}_i=\{{\cal M}_4,g^{(i)}_{\mu\nu}({\bf x}),\phi_i({\bf x})\},\;\;i=1,...,N,\label{gauge-i}\eea where $N\rightarrow\infty$ and each $\phi_i$ belongs in the set of real-valued, smooth continuous functions, provides a potentially different description of the universe. The different gauges belong in the same class of conformal equivalence since appropriate conformal transformations relate one given gauge with any other one in the class. Actually, any two gauges ${\cal G}_i$ and ${\cal G}_j$ are joined by a gauge transformation, 

\bea &&g_{\mu\nu}^{(i)}\rightarrow\Omega^2_{ij}g_{\mu\nu}^{(j)},\;\phi_i\rightarrow\Omega^{-1}_{ij}\phi_j,\nonumber\\
&&\Leftrightarrow\;g_{\mu\nu}^{(i)}\rightarrow\left[\frac{\phi_j}{\phi_i}\right]^2g_{\mu\nu}^{(j)}.\nonumber\eea This means that all gauges in our theory belong in a class of conformal equivalence.

Despite obvious differences, the resulting geometrical picture reminds us the ``many-worlds'' interpretation of quantum physics \cite{everett, dewitt, kent, barvinsky, omnes, tegmark, garriga, zurek, tegmark-nature, page} since different gauges represent different theories, yielding different complete descriptions of the gravitational laws. Below we shall illustrate this in the cosmological context.


\subsection{Many worlds and cosmology}\label{subsect-cosmo-example}

In order to illustrate the geometrical picture outlined above, for definiteness and also for simplicity, let us consider the cosmological setup where the background spacetime is FRW with flat spatial sections. In this case the different gauges can be defined as it follows,

\bea {\cal G}_i=\{{\cal M}_4,a_i^2(\tau),\phi_i(\tau)\},\;\;i=1,...,N,\label{i-gauges}\eea where $N\rightarrow\infty$, $a_i=a_i(\tau)$ is the scale factor in the $i$-th gauge and $\tau$ is the conformal time, which is related with the cosmic time $t$ through, $\tau=\int a^{-1}dt$. Here we prefer to write the relevant quantities as functions of the conformal time instead of the cosmic time, because $\tau$ is not transformed by the gauge transformations \eqref{gauge-t-theor}. Hence, $\tau$ is the same variable in any gauge.

A outstanding gauge is the one generated by the choice $\phi=$const. It is called as GR gauge and to all purposes it is no more than standard general relativity in Riemannian space $V_4$ (see Sect. \ref{subsect-gr-gauge}). It is usually argued that GR is not a gauge invariant theory (it is not). Notwithstanding, in the present theoretical framework, general relativity is no more than a specific gauge of the overall gauge invariant theory described by equations \eqref{wig-tot-lag}, \eqref{eom-theor}, \eqref{cons-theor}, so that manifest gauge invariance is broken down by the gauge choice. Consequently, although gauge symmetry is not a manifest symmetry of general relativity, gauge invariance of the overall theory is implicitly shared by GR as well. The elements of the GR gauge can be expressed as,

\bea {\cal G}_{0k}=\{{\cal M}_4,a_0^2(\tau),\phi_{0k}\},\;\;k\in\mathbb N.\label{i0-gr-gauge}\eea The different constants $\phi_{0k}\in\mathbb R$ generate different sets of physical constants: $\{M^2_{\text{pl},k},\,v_k,...\}$, where $M^2_{\text{pl},k}=\phi^2_{0k}/6$ is the Planck mass squared in the $k$-th element of the GR gauge, $v_k=\phi_{0k}$ is the corresponding mass parameter of the SMP and the ellipsis stand for other possible physical constants which are transformed by the conformal transformations of the metric, such as, for instance, the effective cosmological constant $\Lambda^k_\text{eff}=\Lambda\,\phi^2_{0k}$. Constants such as $\hbar$, the electron charge $e$, the speed of light in vacuum $c$ and the fine structure constant $\alpha$, which are not affected by the gauge transformations \eqref{gauge-t-theor}, are the same in every GR gauge ${\cal G}_{0k}$.

In order to have a more clear idea of the many-worlds picture associated with the existence of infinitely many different gauges in our present theory, let us imagine that, at some initial post-Planckian time 

\bea \tau_0>\tau_\text{pl},\;\tau_\text{pl}\approx\frac{t_\text{pl}}{a(t_\text{pl})},\nonumber\eea where the quantum gravitational effects are subdominant,\footnote{Our present theory with EOMs \eqref{eom-theor} and \eqref{cons-theor} driving the dynamics of gravity in $\tilde W^\text{int}_4$ spaces, is obviously a classical theory of gravity so that we do not expect it to hold true in the quantum domain.} a large number $N$ ($N\rightarrow\infty$) of identical copies of a FRW-$\tilde W^\text{int}_4$ universe that is governed by Eqs. \eqref{eom-theor} and \eqref{cons-theor} (see Eqs. \eqref{fried-eq} and \eqref{cons-eq} below), have been prepared with the same initial conditions. These copies share same particle content, same non-gravitational laws, etc. They differ only in one function: $\phi=\phi(\tau)$ -- the gauge scalar, so that each copy is actually a gauge. Once the $N$ copies are prepared in this initial state, each one of them evolves according to the laws Eqs. \eqref{eom-theor}, \eqref{cons-theor} with the specified functional form of the gauge scalar. After the conformal time $\Delta\tau=\tau-\tau_0$ has elapsed, each one of the copies, say the $i$-th copy, distinguishes from each other. Hence, what we have is a set of $N$ different Universes evolving according to the same laws of gravity \eqref{eom-theor} and \eqref{cons-theor} but with different functions $\phi_i(\tau)$ ($i=1,...,N$). The different gauges represent physically equivalent descriptions of the gravitational laws since these are the same in any gauge. 

Since the number of gauges $N$ can be very large ($N\rightarrow\infty$), in principle any possible pattern of cosmic evolution can be reproduced by given gauges in the set. Hence, the question is: which would be the predictive power of such a theory? In the next section we shall answer this question. As we shall see different gauges fit differently the same observational evidence. The right question then turns out to be which copy -- in the very large number $N$ of copies of the universe -- is the one that better fits the existing amount of observational evidence? The winning copy will be the one where we belong in.


\section{Gauges and observational evidence}\label{sect-observ}

Here we shall look for specific evolution in conformal time of the unknowns $a(\tau)$ and $\vphi(\tau)$, in order to differentiate given gauges. Then we shall seek for observational data fitting of these gauges. As before we shall consider only the high redshift SN-Ia data from Ref. \cite{perlmutter-1999}. Besides, for simplicity, we shall assume vanishing vacuum energy density: $T^\text{vac}_{\nu\mu}=0$ in Eq. \eqref{vac-set}, i. e., we shall assume vanishing effective cosmological constant $\Lambda_\text{eff}=0$. This would entail, in particular, that the GR gauge (basically standard general relativity) can not fit the observational evidence on high redshift SN-Ia.

Let us write the $(00)$-component of the equations of motion \eqref{eom-theor} in terms of the FRW metric \eqref{frw-met},

\bea \frac{3}{a^2}\left(\frac{a'}{a}+\frac{\vphi'}{2}\right)^2=\frac{e^{-\vphi}}{M^2_\text{pl}}\rho_\chi,\label{fried-eq}\eea as well as the null-component of the conservation equation \eqref{cons-theor},

\bea \rho'_\chi+3\frac{a'}{a}\left(\rho_\chi+p_\chi\right)-\frac{\vphi'}{2}\left(\rho_\chi-3p_\chi\right)=0,\label{cons-eq}\eea where the prime accounts for derivative with respect to the conformal time $\tau$. In the above equations we have considered the time normalization where $\vphi(0)=0$ $\Rightarrow\phi(0)=1$, so that the square Planck mass $M^2_\text{pl}=1/6$. We can return to right units at any time by making the following replacement $\phi(0)=\sqrt{6}M_\text{pl}$. Besides, in Eq. \eqref{eom-theor} the stress-energy tensor is that of a prefect fluid, which is given by: 

\bea T^{(\chi)}_{\mu\nu}=(\rho_\chi+p_\chi)u_\mu u_\nu+p_\chi g_{\mu\nu},\nonumber\eea where $\rho_\chi$ and $p_\chi$ are the energy density and the pressure of the perfect fluid, respectively, while $u_\mu=\delta^0_\mu$ is the four-velocity vector of the fluid in the co-moving frame. Let us further, for simplicity, consider pressureless dust ($p_\chi=0$.) In this case the conservation equation \eqref{cons-eq} greatly simplifies. Its integration in quadratures leads to

\bea \rho_\chi=\frac{M^4e^{\vphi/2}}{a^3},\label{rhom-sol}\eea where $M$ is an integration constant with dimensions of mass (do not confound with the absolute magnitude $M$ of distant type IA supernovae.) If we substitute $\rho_\chi$ from \eqref{rhom-sol} back into \eqref{fried-eq}, then the latter equation can be written in the following way: 

\bea e^{\xi}\xi'^2=\frac{M^4}{3M^2_\text{pl}},\label{xi-eq}\eea where we have also introduced the new gauge invariant variable $\xi\equiv\ln a+\vphi/2$. Straightforward integration of Eq. \eqref{xi-eq} yields,

\bea ae^{\vphi/2}=\frac{M^4}{12M^2_\text{pl}}(\tau-\tau_0)^2,\label{avphi-sol}\eea where $\tau_0$ is another integration constant. This solution, as any other solution of the equations of motion \eqref{eom-theor}, \eqref{cons-theor}, shows that we can determine, at most, the gauge invariant combination $\xi=\ln a+\vphi/2$. As already mentioned, this is a direct consequence of gauge invariance: we can choose any $\vphi(\tau)$ we want or, in its place, we can choose any $a=a(\tau)$. The different choices define different gauges. Each gauge amounts to a possible (complete) history of the Universe. The Universe we observe is appropriately described by one of the possible gauges.

Let us consider, for illustrative purposes, three relevant gauges among the infinite set of them:

\begin{enumerate}


\item{\bf GR gauge.} In this case $\vphi=$const. For definiteness we choose $\vphi=0$. This is one of the elements in the infinite set of elements in the GR gauge (recall that other values of the constant $\vphi=\vphi_0$ also define possible elements of the GR gauge.) In this case Eq. \eqref{avphi-sol} simplifies to,\footnote{Here we replace a hat over a quantity by the subindex ''GR'' to denote a quantity in the GR gauge over Riemann space $V_4$.}

\bea a_\text{GR}(\tau)=\frac{M^4}{12M^2_\text{pl}}\left(\tau-\tau_0\right)^2,\label{gr-gauge-1}\eea where $\tau_0$ marks the starting point of the cosmic expansion and we shall use the following conformal time normalization: at present time $\tau=\tau_*$ we have that $a_\text{GR}(\tau_*)=1$. Hence, from Eq. \eqref{gr-gauge-1} it follows that,

\bea \tau_*-\tau_0=\sqrt\frac{12M^2_\text{pl}}{M^4}.\label{t-aster}\eea Besides (recall that $dt=ad\tau$),

\bea H_\text{GR}=\frac{\dot a_\text{GR}}{a_\text{GR}}=\frac{a'_\text{GR}}{a_\text{GR}^2}=\frac{24M^2_\text{pl}}{M^4}\left(\tau-\tau_0\right)^{-3}.\label{gr-gauge-2}\eea We have also,

\bea \dot H_\text{GR}=\frac{H'_\text{GR}}{a_\text{GR}}=-\frac{864M^4_\text{pl}}{M^8}\left(\tau-\tau_0\right)^{-6}.\label{gr-gauge-3}\eea Hence, the deceleration parameter,

\bea q\equiv-\left(1+\frac{\dot H_\text{GR}}{H_\text{GR}^2}\right)=\frac{1}{2},\label{gr-gauge-4}\eea so that the expansion is decelerated.

Taking into account \eqref{avphi-sol} and that $ae^{\vphi/2}=a_\text{GR}=(1+z_\text{GR})^{-1}$, we can write the conformal time in terms of the redshift $z_\text{GR}$,

\bea \tau-\tau_0=\frac{\sqrt{12M^2_\text{pl}/M^4}}{\sqrt{1+z_\text{GR}}}.\label{tau-z}\eea For the comoving distance we have that,

\bea d_\text{C}=\int\frac{da}{a^2H(a)}=\int d\tau=\tau-\tau_0,\label{comov-chi}\eea i. e., modulo a constant, the comoving distance coincides with the conformal time. By deriving \eqref{tau-z} we get that,

\bea d\tau=-\sqrt\frac{3M^2_\text{pl}}{M^4}\frac{dz_\text{GR}}{\left(1+z_\text{GR}\right)^{3/2}},\nonumber\eea which, after integration, yields

\bea d_\text{C}=\tau=\sqrt\frac{3M^2_\text{pl}}{M^4}\int_0^{z_\text{GR}}\frac{dz'_\text{GR}}{\left(1+z'_\text{GR}\right)^{3/2}}.\label{com-dist-zgr}\eea Comparing this equation with Eq. \eqref{h0dl} with $\Omega^0_m=1$ and $\Omega^0_\Lambda=0$, and comparing with Eq. \eqref{dl}, one gets that the integration constant,

\bea M^4=3H^2_0M^2_\text{pl},\label{m4}\eea where $H_0$ is the present value of the GR Hubble parameter. Actually, substituting Eq. \eqref{m4} into Eq. \eqref{tau-z} yields that Eq. \eqref{gr-gauge-2} can be written in the following way:

\bea H_\text{GR}\left(z_\text{GR}\right)=H_0\left(1+z_\text{GR}\right)^{3/2},\eea so that $H_\text{GR}(0)=H_0$.

The plot of the apparent magnitude vs redshift $z_\text{GR}$ for this gauge: general relativity with a vanishing cosmological constant, corresponds to the solid curve in the left panel of FIG. \ref{fig1}. It is seen that, in what regards to the high-redshift type IA supernovae data, this gauge can not explain the resulting accelerated pace of the cosmic expansion.


\item{\bf Flat gauge.} In this case Minkowski background metric $\eta_{\mu\nu}$ is assumed. Since the curvature of spacetime vanishes, then $a=a_0=1$, which if substituted in Eq. \eqref{avphi-sol} yields

\bea e^{\vphi/2}=\frac{M^4}{12M^2_\text{pl}}\left(\tau-\tau_0\right)^2.\label{flat-sol}\eea In the present case the comoving distance obeys Eq. \eqref{comov-chi}, and the only contribution to the redshift comes from variation of masses due to gradient nonmetricity. Hence,

\bea e^{\vphi/2}=\frac{1}{1+z_\text{GR}}.\label{vphi-zgr}\eea Combining Eqs. \eqref{flat-sol}, \eqref{vphi-zgr} and \eqref{comov-chi}, for the comoving distance one gets the same expression \eqref{com-dist-zgr} that we obtained in the former gauge. What this means is that, although in the present gauge the universe is not expanding $H=0$, the fit of the model to the observational data on high-redshift type IA supernovae is the same as in the GR gauge with vanishing cosmological constant $\Lambda_\text{eff}=0$ (solid curve in the left figure in FIG. \ref{fig1}). This means that the model is ruled out by the cosmological observations.\footnote{Only if add the problematic cosmological constant term in the above and in the present gauges we obtain a good fit to the data on high-redshift measurements (dash-dot curve in the left panel of FIG. \ref{fig1}).}

In addition, the flat gauge is phenomenologically ruled out because the light rays follow straight lines and do not suffer gravitational bending. Actually, since, on the one hand, photons do not interact with nonmetricity, in flat background space the null geodesics Eq. \eqref{null-auto-p}, read

\bea \frac{dk^\alpha}{d\lambda}=0.\nonumber\eea On the other hand, in the flat gauge gravity is a manifestation of nonmetricity exclusively. Therefore, photons do not suffer neither curvature redshift of frequency nor gravitational bending. Means that this model does not pass the Solar system test on light bending.


\item{\bf Third gauge.} Let us consider a third (``intermediate'') possibility where neither the scale factor $a$, nor the gauge scalar $\vphi$ are constants. Since the product $a\exp(\vphi/2)$ is gauge invariant, this means that $a\exp(\vphi/2)=a_\text{GR}=1/\left(1+z_\text{GR}\right)$, where $a_\text{GR}$ is the scale factor in the GR gauge. Hence, according to Eq. \eqref{avphi-sol} the following relationship between the conformal time $\tau$ and the redshift $z$, can be established:

\bea 1+z_\text{GR}=\frac{4}{H^2_0\left(\tau-\tau_0\right)^2},\label{tau-z'}\eea where we have considered Eq. \eqref{m4}. In this case the redshift of frequency is contributed both by the curvature of space and by nonmetricity. The overall redshift $z_\text{tot}$ is given by Eq. \eqref{z-tot}, which can be written as in Eq. \eqref{z-tot-fin}. In Sect. \ref{sect-z-q} we have shown that an acceptable fit to the high-redshift data set of Ref. \cite{perlmutter-1999} is obtained if assume that $\vphi=\epsilon z_\text{GR}$ is a small quantity, so that the overall redshift can be written in terms of the GR redshift through Eq. \eqref{z-master}. Here we shall assume this is the case, so that the ``third gauge'' is consistent with the observations. Of course the above is only an approximate expression for the gauge scalar $\vphi$, which is valid for redshifts $z_\text{GR}<1$. This means that the correct gauge should contain this (or a similar) approximation as a particular limit. Yet, since our discussion is mostly illustrative (qualitative), it suffices to consider the above approximation as a gauge covering the whole cosmic history from the distant past $z_\text{GR}\rightarrow\infty$ to the asymptotic future $z_\text{GR}\rightarrow-1$.

In this gauge, from Eq. \eqref{avphi-sol} it follows that,

\bea a(\tau)=\frac{e^{\epsilon(1-\zeta)/2}}{\zeta},\nonumber\eea where we have defined the following function of the conformal time $\tau$:

\bea \zeta=\zeta(\tau)\equiv\frac{4}{H^2_0\left(\tau-\tau_0\right)^2}.\nonumber\eea Besides, the Hubble parameter as function of the conformal time reads

\bea H(\tau)=H_0\zeta^{3/2}e^{\epsilon(\zeta-1)/2}\left(1+\frac{\epsilon}{2}\zeta\right).\nonumber\eea Given our normalization in Eqs. \eqref{t-aster} and \eqref{m4}, it follows that at present time $\tau=\tau_*$, $\tau_*-\tau_0=2H_0^{-1}$, so that $\zeta(\tau_*)=1$. In consequence in this gauge $a(\tau_*)=1$, which leads the present value of the Hubble parameter,

\bea H(\tau_*)=H_0\left(1+\frac{\epsilon}{2}\right),\label{hubble-present}\eea to slightly differ from its GR value $H_\text{GR}(\tau_*)=H_0$. In general, for arbitrary function $\vphi=\vphi(z_\text{GR})$, we have that,

\bea H=H_0e^\frac{\vphi}{2}\left(1+z_\text{GR}\right)^\frac{3}{2}\left(1+\frac{1+z_\text{GR}}{2}\frac{d\vphi}{dz_\text{GR}}\right),\label{h-arb-vphi}\eea while the overall redshift and the GR redshift are related by:

\bea z_\text{tot}+2=e^{\vphi/2}\left(z_\text{GR}+2\right).\nonumber\eea


\end{enumerate}

The question under scrutiny is, which would be the predictive power of a theory that admits almost any possible evolution pattern? In order to answer this question let us first summarize our above results.

Among the infinite number of different gauges, above we have chosen three of them: i) the GR gauge, ii) the flat gauge and iii) the third gauge. The different gauges provide different but equivalent descriptions of the cosmological evolution. While in the GR gauge gravity is due to the curvature of FRW-$V_4$ (Riemann) space, in the flat space gauge it is completely due to gradient nonmetricity in Minkowski space. In the third gauge the gravitational phenomena are associated both with curvature of FRW-$\tilde W^\text{int}_4$ space and with gradient nonmetricity. The different gauges are physically equivalent since they satisfy the same laws of gravity Eqs. \eqref{eom-theor}, \eqref{cons-theor}. Gauge equivalence is due to invariance of the gauge-invariant quantities such as, for instance $a\exp(\vphi/2)$, under the transformations \eqref{gauge-t-theor}. 


The critical argument in order to answer the question on the predictive power of our theory is the following. Although all three gauges above: GR, flat and third gauges, yield different but equivalent descriptions of the same gravitational laws, only one of them: the third gauge, fits well the observational evidence from high-redshift type IA supernovae. Neither the GR gauge (vanishing cosmological constant) nor the flat one fit well enough the observational SN-Ia data. In addition, the flat gauge is not compatible with light bending in a gravitational field so that Solar system tests rule out this gauge. In consequence, the GR and flat gauges are ruled out by experiments/observations. 

Experiment and observations in general, play a crucial role in determining which one, in the large number $N\rightarrow\infty$ of possible gauges, is the one that better describes our causally accessible universe. In a sense experiment allows us to determine the gauge ``which we live in,'' which in what follows we shall call as ``world-gauge.'' Once this gauge is fully determined, which means that we fully determined the gauge scalar $\vphi=\vphi({\bf x})$, one can make predictions with the help of the gravitational laws \eqref{fried-eq} and \eqref{cons-eq}, which are valid in our world-gauge.


\section{The $H_0$ tension issue}\label{sect-hubble-t}


According to the discussion in the former section, observations and experiments favor the third gauge, which we identify with the correct description of our causally accessible universe. One of the predictions we can make on the basis of the laws governing our world-gauge, is that the present value of the Hubble parameter $H(\tau_*)$ in Eq. \eqref{hubble-present} differs from the one computed on the basis of the GR equations ($H_0$). This discrepancy is unavoidable. Actually, working with equation \eqref{avphi-sol}, with due consideration of equations \eqref{t-aster} and \eqref{m4}, one obtains the expression for the Hubble parameter as a function of the conformal time in the most general case:

\bea \frac{H(\tau)}{H_0}=e^\frac{\vphi}{2}\left(\frac{\tau_*-\tau_0}{\tau-\tau_0}\right)^3\left[1-\left(\frac{\tau-\tau_0}{\tau_*-\tau_0}\right)\frac{\vphi'}{2H_0}\right],\label{hubble-tau-gen}\eea where both $\vphi=\vphi(\tau)$ and $\vphi'=\vphi'(\tau)$ are functions of the conformal time. We have that, at present time, $\vphi_*=\vphi(\tau_*)=0$ and $\vphi'_*=\vphi'(\tau_*)=-H_0s_*\neq 0$, respectively ($s_*$ is some constant.) In this equation $H_\text{GR}=H_0$ is the today value of the Hubble constant computed in the GR gauge so that, if evaluate Eq. \eqref{hubble-tau-gen} at present time $\tau=\tau_*$, one gets

\bea H(\tau_*)=H_\text{GR}\left(1+\frac{s_*}{2}\right).\label{hubble-tension}\eea Hence, if the gravitational laws which describe our Universe differ from the GR ones, i. e., if the gauge we live in is not the GR gauge, the above discussed discrepancy is unavoidable.

This discrepancy may explain the disagreement between the present value of the Hubble parameter locally measured (no specific model assumed) of about $H_0\approx 73.2$ km s$^{-1}$ Mpc$^{-1}$ at 68 $\%$ confidence level and the one inferred from GR (plus the cosmological constant) of about $H_0\approx 67.3$ km s$^{-1}$ Mpc$^{-1}$ by evaluating early times physics \cite{divalentino-rev, biaggio-2021, biaggio-2022}. Actually, under the assumption that at redshifts $z<1$ our approximation \eqref{z-master} is not bad, if in equation \eqref{hubble-present} set $\epsilon\approx 0.15$, or without specific assumptions if in \eqref{hubble-tension} set $s_*\approx 0.15$, the mentioned discrepancy is explained: Assuming that the present value of the Hubble parameter computed in our world-gauge coincides with the one measured in local experiments $H(\tau_*)\approx 73.2$ km s$^{-1}$ Mpc$^{-1}$, from \eqref{hubble-tension} one gets that the corresponding GR value $H_\text{GR}\approx 68.1$ km s$^{-1}$ Mpc$^{-1}$. These estimates can be improved by improving our knowledge of the gauge scalar as function of the conformal time $\vphi(\tau)$. 

According to the above explanation the $H_0$ tension issue \cite{divalentino-rev} is due to the assumption of an incorrect theory: general relativity, in particular the $\Lambda$CDM model, in order to compute the present value of the Hubble constant by evaluating early time physics. In our gauge invariant framework we do not need neither GR nor the cosmological constant in order to explain the accelerated expansion of the cosmos (see Sec. \ref{sect-z-q}). Hence, if assume that the theory \eqref{fried-eq}, \eqref{cons-eq} correctly describes the classical laws of gravity, including the early times stage, there will be no discrepancy between the present value of the Hubble constant computed by extrapolating early times physics to the present, with the one inferred from local measurements (late time physics).



\section{Flatness, horizon and relict particles abundance problems}\label{sect-flatness}


Any theoretical framework that pretends to explain the past, present and future of the cosmic expansion, should be able to explain the flatness, horizon and relict particles abundance problems. Cosmic inflation \cite{infl-guth, infl-albrecht, infl-linde-1, infl-linde-2, olive-phys-rept-1990, hyb-infl-linde, infl-reheat, liddle-prd-1994, barrow-prd-1995, lidsey-rmp-1997, assist-infl, infl-lyth, liddle-book, infl-n-gauss} has been developed, precisely, with the aim to explain these puzzles. Here we shall show that the neither the flatness nor the horizon and relict particle abundance problems arise in the present gauge invariant theory, so that no inflationary mechanism is required.


\subsection{Flatness problem}

Let us briefly explain what is the flatness problem about. For this purpose let us to write the GR equations of motion Eqs. \eqref{eom-v4}, \eqref{cons-v4} in FRW background with the following line element in comoving spherical coordinates $t,r,\theta,\phi$:

\bea ds^2_\text{GR}=-dt^2_\text{GR}+\frac{a^2_\text{GR}}{1-kr^2}dr^2+r^2a^2_\text{GR}d\Omega^2,\label{frw-k}\eea where $a_\text{GR}$ is the scale factor, $k=\pm 1,0$ account for the curvature of the spatial sections and $d\Omega^2\equiv d\theta^2+\sin^2\theta d\phi^2$. The FRW GR-EOM read:

\bea &&3H^2_{GR}+\frac{3k}{a^2_{GR}}=\frac{1}{M^2_\text{pl}}\rho_\chi,\label{fried-gr}\\
&&\dot\rho_\chi+3\gamma_\chi H_{GR}\rho_\chi=0,\label{cont-frw}\eea where $H_{GR}=\dot a_{GR}/a_{GR}$ is the GR Hubble rate, $M_\text{pl}$ is the Planck mass and $\rho_\chi$, $p_\chi=(\gamma_\chi-1)\rho_\chi$ are the energy density and the barotropic pressure of the background matter fluid, respectively ($\gamma_\chi$ is the barotropic index of the matter fluid.) In terms of the dimensionless (normalized) energy density: $\Omega_\chi\equiv\rho_\chi/3M^2_\text{pl}H^2_{GR},$ the Friedmann equation \eqref{fried-gr} can be written in the following alternative way: 

\bea |\eta|:=|\Omega_\chi-1|=\frac{|k|}{a^2_{GR} H^2_{GR}},\label{ratio-gr}\eea where the quantity $\eta$ measures the departure from spatial flatness. Integration of Eq. \eqref{cont-frw} yields $\rho_\chi=M^4a^{-3\gamma_m}$ ($M$ is an integration constant with mass unit,) so that \eqref{ratio-gr} can be written in the following way:

\bea |\eta|=\frac{|k|}{\left|H^2_0\,a^{2-3\gamma_\chi}-k\right|},\label{xi-gr}\eea where we took into account \eqref{m4}. From this equation it follows that for $\gamma_\chi>2/3$, a requirement that is satisfied by dust ($\gamma_\chi=1>2/3$) and also by radiation ($\gamma_\chi=4/3>2/3$), the cosmic expansion leads to $|\eta|\rightarrow 1$ at late times. 

The only way in which the departure from spatial flatness decreases with the curse of the cosmic expansion is for matter with $\gamma_\chi<2/3$. In this case, for $a_{GR}\gg(3kM^2_\text{pl}/M^4)^{1/(2-3\gamma_\chi)}$,

\bea |\eta|\approx\frac{3kM^2_\text{pl}}{M^4a_{GR}^{2-3\gamma_\chi}},\nonumber\eea so that, at late times $|\eta|\rightarrow 0$. The inflaton $\sigma$ is a kind of dynamical self-interacting scalar field which can lead to fulfillment of the condition $\gamma_\sigma<2/3$. The latter requirement is obviously satisfied by vacuum fluid ($\gamma=0$) as well.

In order to show that this problem does not arise in our gauge invariant theory, let us write the EOM \eqref{eom-theor} and \eqref{cons-theor} in terms of the metric \eqref{frw-k} written in an arbitrary gauge:

\bea ds^2=-dt^2+\frac{a^2}{1-kr^2}dr^2+r^2a^2d\Omega^2.\label{frw-k'}\eea The resulting EOMs read:

\bea &&3\left(H+\frac{\dot\vphi}{2}\right)^2+\frac{3k}{a^2}=\frac{e^{-\vphi}}{M^2_\text{pl}}\rho_\chi,\label{frw-k-theor}\\
&&\dot\rho_\chi+\left(3H-\frac{\dot\vphi}{2}\right)\rho_\chi=0,\label{cons-k-theor}\eea where, for simplicity, we have chosen dust fluid ($p_\chi=0$). Let us for definiteness consider open Universe ($k=-1$) exclusively. Taking into account the gauge invariant quantity $\xi=ae^{\vphi/2}=a_\text{GR}$, one finds that

\bea &&a^2_\text{GR}H^2_\text{GR}=a^2e^\vphi\left(H+\frac{\dot\vphi}{2}\right)^2\nonumber\\
&&\;\;\;\;\;\;\;\;\;\;\;\;\;\;\;=a^2e^\vphi\left(\frac{e^{-\vphi}\rho_\chi}{3M^2_\text{pl}}+\frac{1}{a^2}\right).\nonumber\eea Hence Eq. \eqref{ratio-gr} can be written in terms of variables of an arbitrary gauge:

\bea |\eta|=\frac{e^{-\vphi}}{\left|\frac{a^2e^{-\vphi}\rho_\chi}{3M^2_\text{pl}}+1\right|}.\label{gen-flat}\eea Integration of the conservation equation \eqref{cons-k-theor} yields $\rho_\chi=\rho_0e^{\vphi/2}a^{-3}$, where $\rho_0$ is an integration constant, so that

\bea |\eta|=\frac{e^{-\vphi}|\xi|}{\left|\frac{\rho_0}{3M^2_\text{pl}}+\xi\right|}.\label{dust-flat}\eea The gauge variable $\xi$ can be found as solution of the Friedmann equation \eqref{frw-k-theor},

\bea \xi(\tau)=\frac{\rho_0}{3M^2_\text{pl}}\sinh^2\left[\sqrt{6}M_\text{pl}\left(\frac{\tau-\tau_0}{2}\right)\right],\label{dust-sol}\eea where in the sinh argument we replaced $1\rightarrow\sqrt{6}M_\text{pl}$ in order to meet correct units. Therefore, at $\tau-\tau_0\gg M^{-1}_\text{pl}$,

\bea \xi(\tau)\approx\frac{\rho_0}{12M^2_\text{pl}}e^{\sqrt{6}M_\text{pl}(\tau-\tau_0)}.\nonumber\eea Substituting this into \eqref{dust-flat} we obtain that $|\eta|\approx e^{-\vphi}$. As seen gauge freedom can explain flatness. All we need is that during the very first stages of the cosmic evolution $\vphi=\vphi(\tau)$ was a growing function of the conformal time.


\begin{figure*}[t!]\begin{center}
\includegraphics[width=5cm]{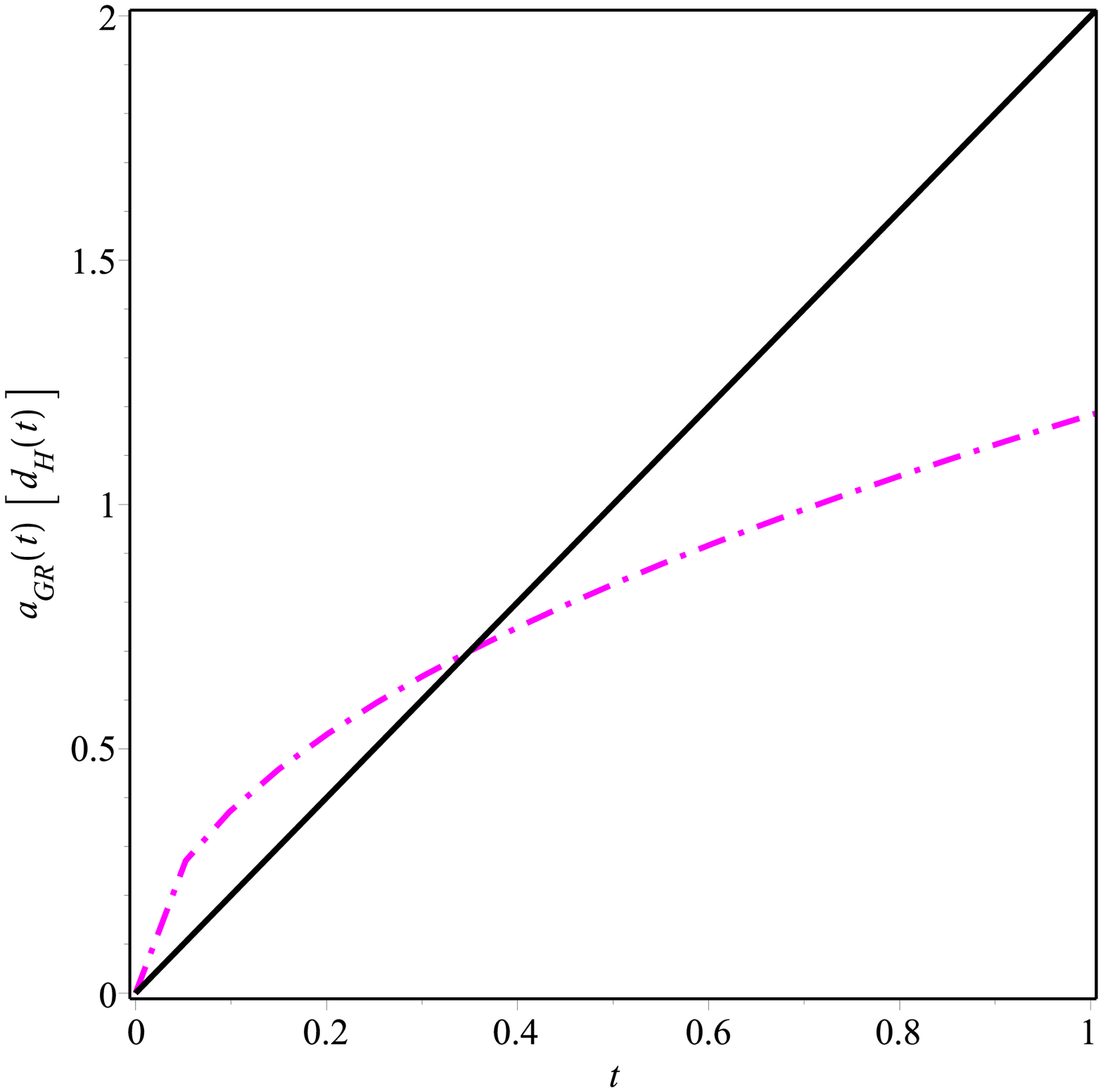}
\includegraphics[width=5cm]{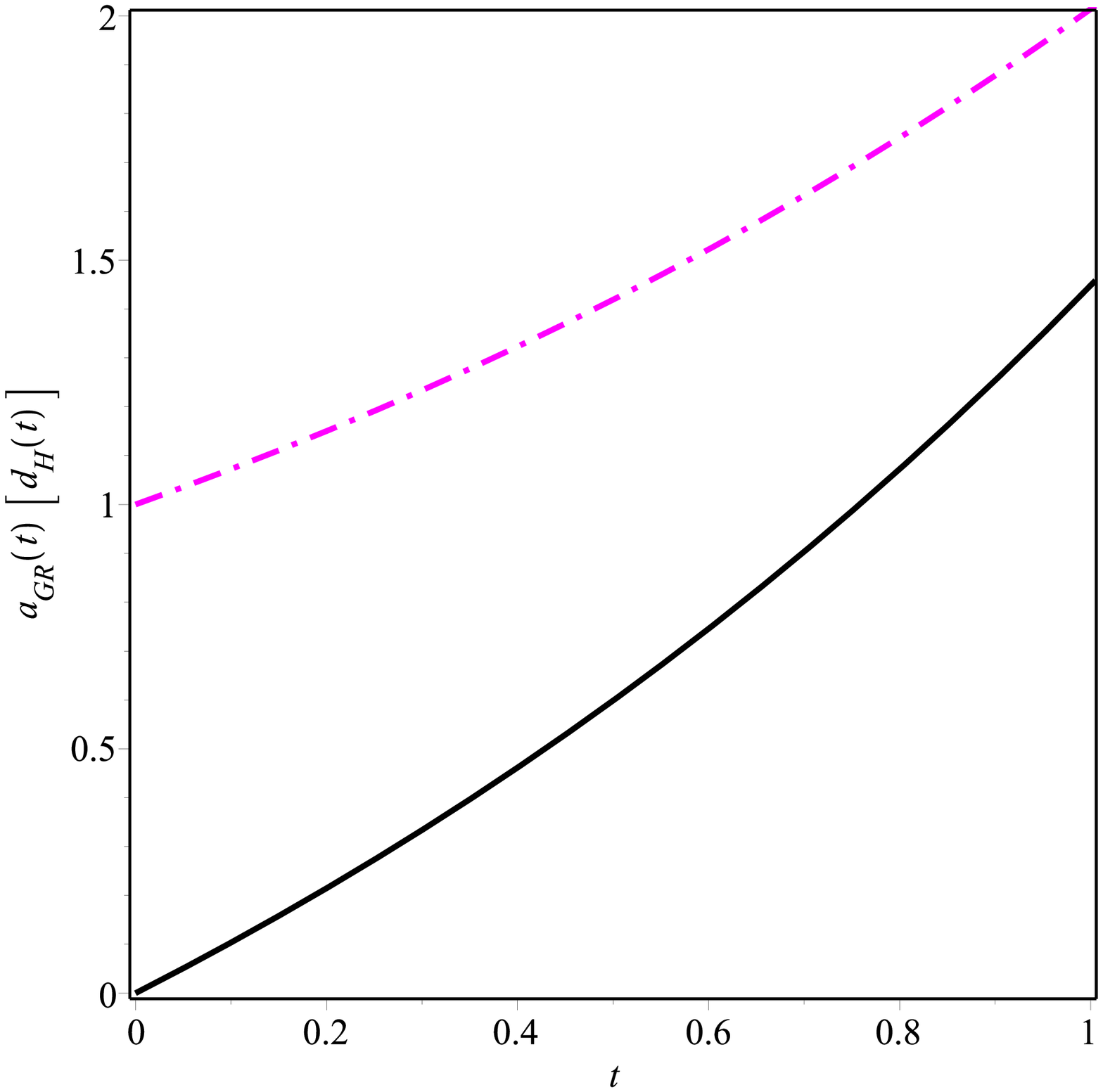}
\includegraphics[width=5cm]{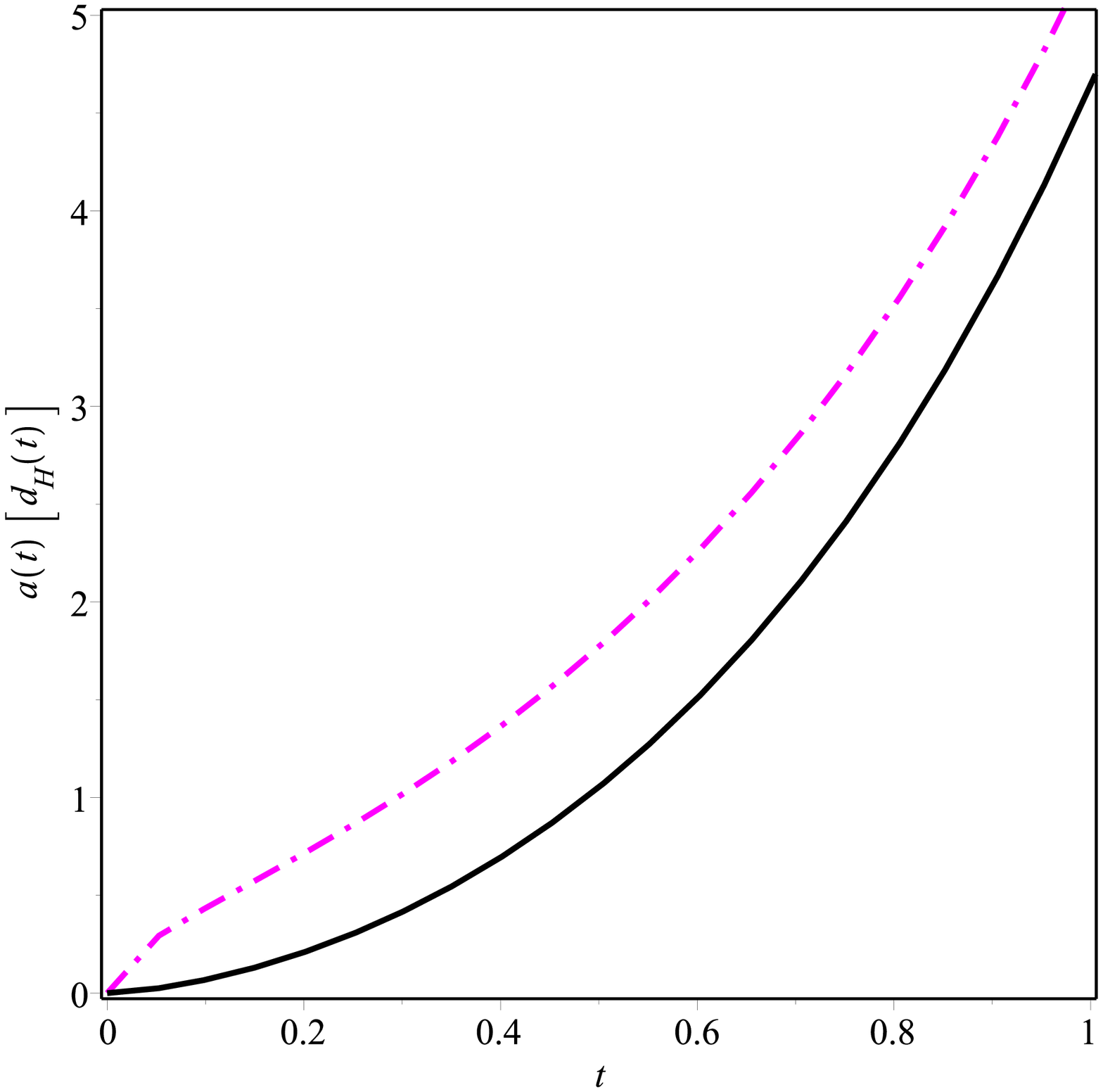}\end{center}
\vspace{1.3cm}\caption{Plots of the scale factor and of the distance to the causal horizon vs cosmic time $t$ (dash-dot and solid curves, respectively). We arbitrarily chose the following values of the constants: $H_0=0.7$, $\mu=1.5$. The left figure is for GR with background radiation where the horizon problem is evident (equations \eqref{gr-scale-f} and \eqref{dh}), while the middle figure depicts GR-de Sitter expansion where the horizon puzzle is settled (equations \eqref{gr-scale-f-infla} and \eqref{dh-infla}). In the right figure the plots are for an arbitrarily chosen gauge of the present gauge invariant theory. This gauge is specified by equations \eqref{scale-f-ginv} for the scale factor and \eqref{dh-ginv} for the distance to the causal horizon. In this last case the horizon issue does not arise.}\label{fig2}\end{figure*}



\subsection{Horizon problem}

In order to simplify the analysis in this subsection we shall consider FRW spacetime with flat spatial sections ($k=0$.) For definiteness let us consider background radiation $p_r=\rho_r/3$. The cosmological equations read,

\bea &&3\left(H+\frac{\dot\vphi}{2}\right)^2=\frac{e^{-\vphi}}{M^2_\text{pl}}\rho_r,\label{fried-hor}\\
&&\dot\rho_r+4H\rho_r=0\;\Rightarrow\;\rho_r=\frac{M^4}{a^4}.\label{cons-rad}\eea

The horizon problem arises because scales that originated outside of the causal horizon will eventually enter our past light cone and hence these will become part of our observable Universe \cite{olive-phys-rept-1990}. In consequence anisotropies are expected to be observed on large scales \cite{rindler, ellis}. Within the GR framework length scales grow as the scale factor:

\bea a_{GR}(t)=\sqrt{2H_0}\,t^{1/2}.\label{gr-scale-f}\eea Meanwhile, the causal horizon \cite{ellis}, which amounts to the maximal physical distance light can travel from the co-moving position of an observer at some initial time to time $t$ \cite{olive-phys-rept-1990}: 

\bea d_H(t)=a_{GR}\int_0^t\frac{dt'}{a_{GR}(t')}=2t.\label{dh}\eea Hence, the distance to the causal horizon grows faster than co-moving separations i. e., than the scale factor. 

The horizon problem is illustrated in the left figure of FIG. \ref{fig2} where the plots of $a_{GR}$ (dash-dots) and of $d_H$ (solid curve) vs the cosmic time $t$, are shown. It is seen that at early times, very close to the bigbang, $t_{bb}$ and up to the ``equality time'' $t_{eq}$: time at which the dash-dot and the solid curves meet again, the curve representing $d_H(t)$ lies below of the curve for $a_{GR}(t)$. During this time interval, scales that at certain $t_{bb}\leq t\leq t_{eq}$ were located above the solid curve and below of the dash-dots, are not in causal contact with the co-moving position, while scales that are located below of the solid curve are causally connected with co-moving observer instead. After the equality time $t_{eq}$ those scales that were out of causal contact since the bigbang $t_{bb}$ and up to $t_{eq}$, enter the causal horizon so these can be seen by a co-moving observer. 

Inflation can take account of the horizon problem since during the de Sitter expansion period: 

\bea a_{GR}(t)=\exp(H_0\,t),\label{gr-scale-f-infla}\eea while 

\bea d_H(t)=\frac{1}{H_0}\left(e^{H_0\,t}-1\right).\label{dh-infla}\eea The plots of $a_{GR}$ Eq. \eqref{gr-scale-f-infla} and of $d_H$ Eq. \eqref{dh-infla} for this case are shown in the middle figure of FIG. \ref{fig2}. It is seen that scales that in the past were out of causal contact keep causally disconnected for all time.

As it was for the flatness problem, within the present gauge invariant framework, due to gauge freedom, inflation is not required in order to explain the horizon issue. Since $\xi=ae^{\vphi/2}$ is a gauge invariant quantity, for the present case: spatially flat FRW metric with background radiation, we have that,

\bea a_{GR}=\sqrt{2H_0}\,\sqrt{t}\;\Rightarrow\;a(t)=\sqrt{2H_0}\,\sqrt{t}\,e^{-\vphi/2},\nonumber\eea where Eq. \eqref{gr-scale-f} has been taken into account. Let us choose the gauge with $\vphi(t)=-2\mu\,\sqrt{t}$, where $\mu$ is a constant parameter with the dimensions of the square of mass. For the scale factor and the distance to the causal horizon we get that,

\bea &&a(t)=\sqrt{2H_0}\,\sqrt{t}\,e^{\mu\sqrt{t}},\label{scale-f-ginv}\\
&&d_H(t)=\frac{2\sqrt{t}}{\mu}\left(e^{\mu\sqrt{t}}-1\right),\label{dh-ginv}\eea respectively. In this gauge, for $\mu\leq\sqrt{2/H_0}$, those scales which were out of causal contact in the past, will be causally disconnected for all future times, as it was for inflation within the GR framework. This is illustrated in the right figure in FIG. \ref{fig2}, where the dash-dot curve represents the evolution of the scale factor \eqref{scale-f-ginv} while the solid curve represents the evolution of the distance to the causal horizon $d_H$ in cosmic time Eq. \eqref{dh-ginv}.


\begin{figure*}[t!]\begin{center}
\includegraphics[width=7.5cm]{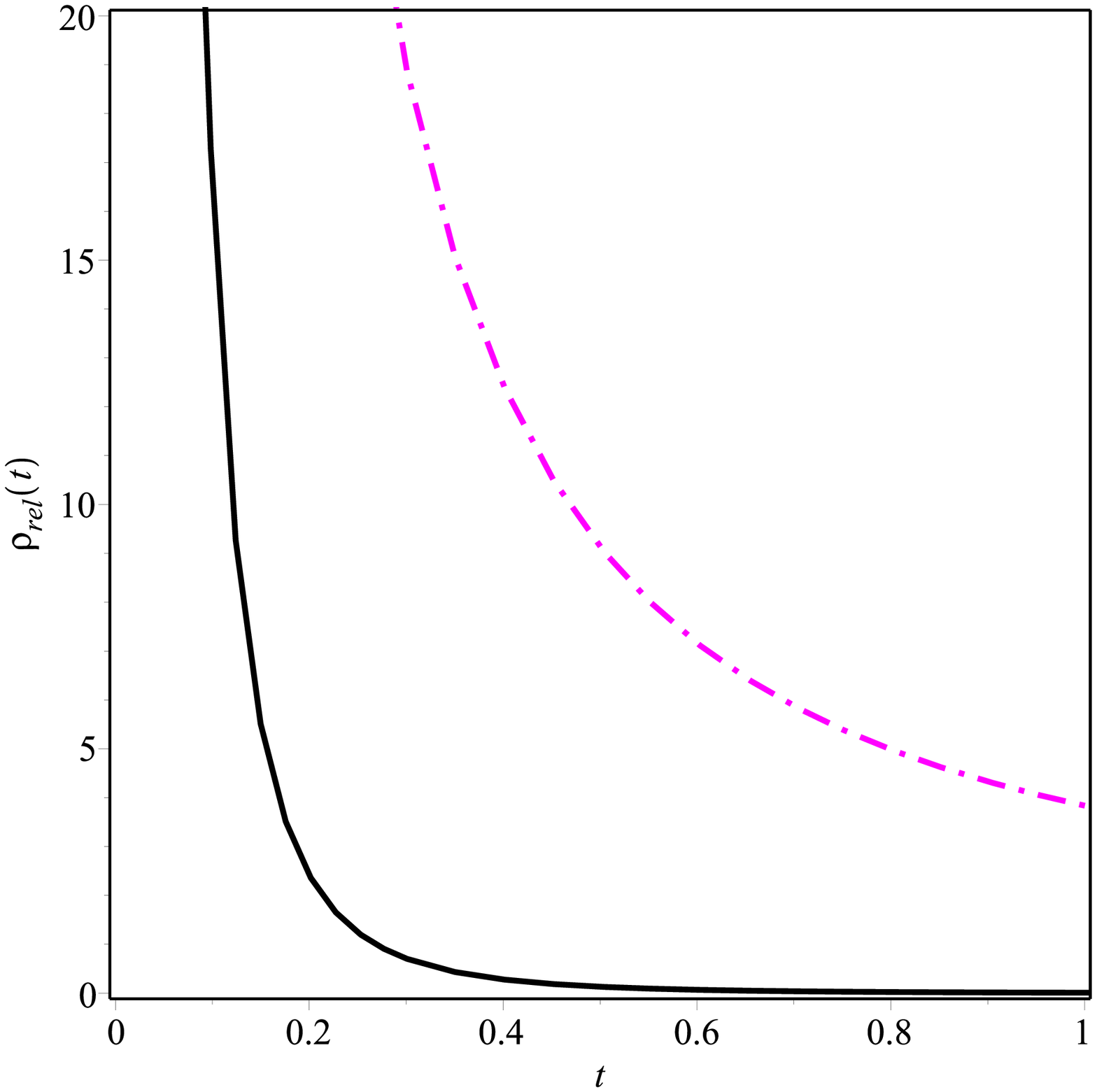}
\includegraphics[width=7.5cm]{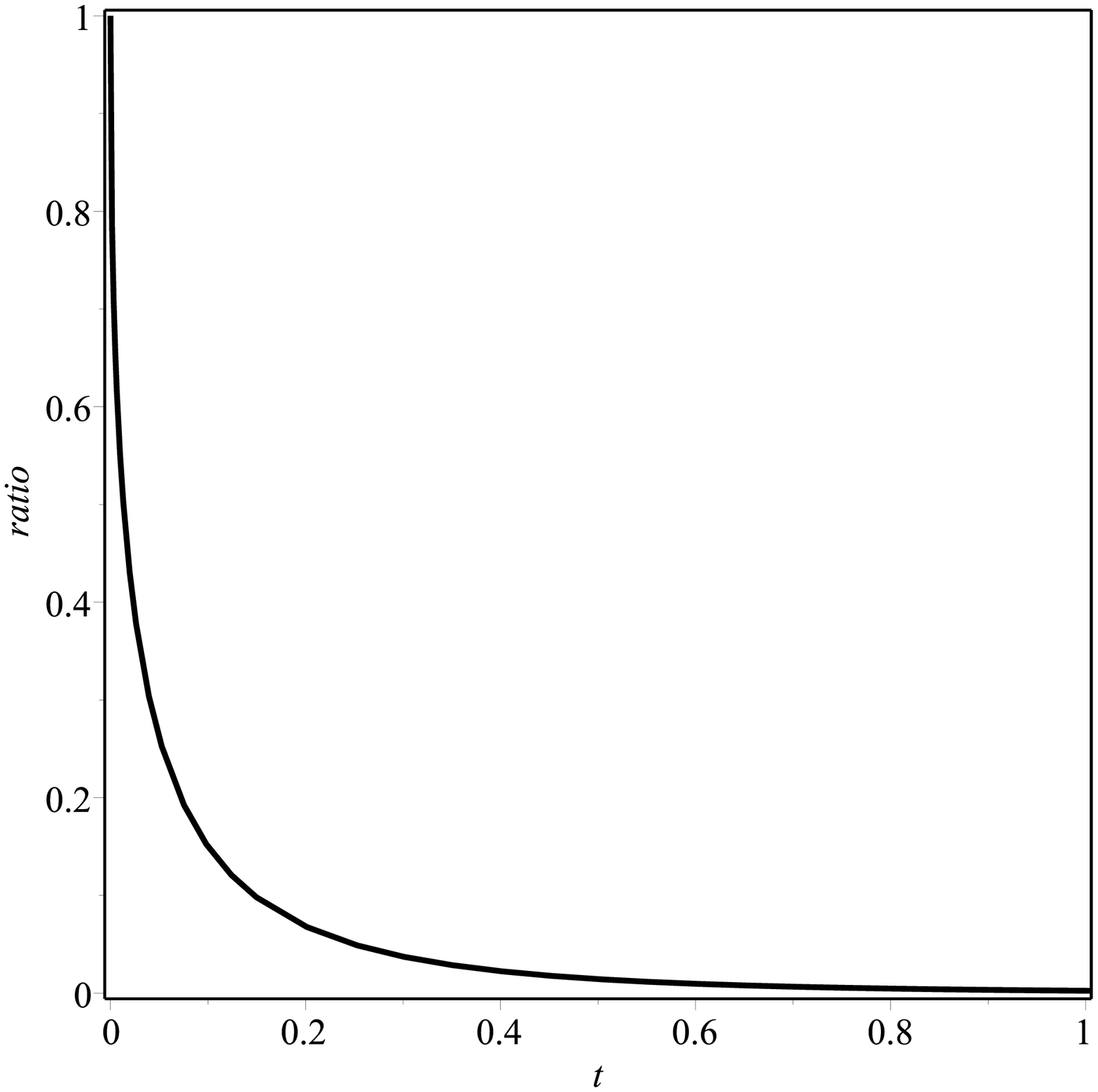}\end{center}
\vspace{1.3cm}\caption{In the left figure the plots of the energy densities $\rho^{GR}_\text{rel}$ in Eq. \eqref{rogr} -- dash-dots -- and $\rho_\text{rel}$ in Eq. \eqref{rorel} -- solid curve -- vs cosmic time $t$ are shown. We arbitrarily chose the following values of the constants: $H_0=0.7$, $\mu=1.5$, $M^2_\text{pl}=1$ and $\gamma=1$ (dust of relict particles). The ratio $\rho_\text{rel}/\rho^{GR}_\text{rel}$ in Eq. \eqref{ratio} vs cosmic time $t$ is drawn in the right figure. It is seen that, independent of the behavior of $\rho^{GR}_\text{rel}(t)$, the energy density of relict particles in the present gauge of our gauge invariant theory, very quickly dilutes with the expansion.}\label{fig3}\end{figure*}



\subsection{Abundance of relict particles}

In a similar fashion gauge freedom may explain the problem with the abundance of relict particles such as magnetic monopoles, gravitinos, moduli fields, etc. The question in this case is why the universe is not dominated by these heavy relict particles at present? In order to explain why this is not so within GR-based cosmology, it is again required the inflationary stage, so that any initially existing amount of relict particles will be very quickly diluted 

\bea \rho^{GR}_\text{rel}\propto\exp(-3\gamma H_0\,t).\nonumber\eea This means, in turn, that within the GR framework an additional inflaton field is required.

In the present gauge invariant theory inflation is not necessary to explain the problem with the abundance of relict particles, as we shall see. In this case the gravitational laws are given by Eqs. \eqref{fried-eq} and \eqref{cons-eq}, respectively. As before, for simplicity we shall consider spatially flat FRW background space. We also assume that the background fluid is a perfect fluid of relict particles with energy density and pressure $\rho_\text{rel}$, $p_\text{rel}$, which satisfy $p_\text{rel}=(\gamma-1)\rho_\text{rel}$, where $\gamma$ is the barotropic index of the relict fluid.\footnote{This analysis is not exact since, in general the equation of state of the fluid of relict particles is very complex so that $p_\text{rel}\neq(\gamma-1)\rho_\text{rel}$.} Integration of \eqref{cons-eq} yields (as before we use the gauge invariant variable $\xi\equiv ae^{\vphi/2}=a_{GR}$,)

\bea \rho_\text{rel}=\frac{M^4\,e^{\frac{4-3\gamma}{2}\vphi}}{a^{3\gamma}}=\frac{M^4\,e^{2\vphi}}{\xi^{3\gamma}}=\frac{M^4\,e^{2\vphi}}{a_{GR}^{3\gamma}},\nonumber\eea or

\bea \rho_\text{rel}=e^{2\vphi}\rho^{GR}_\text{rel},\label{rho-sol}\eea where $\rho^{GR}_\text{rel}$ is the solution of the GR conservation equation $\dot\rho^{GR}_\text{rel}+3\gamma H_{GR}\,\rho^{GR}_\text{rel}=0$. We can find a gauge where the density of the fluid of relict particles very quickly decays with the curse of the cosmic expansion. In particular, the gauge $\vphi(t)=-2\mu\,\sqrt{t}$ which solved the horizon problem above, can take account of the abundance of relict particles as well. In this gauge the Friedmann equation can be written as,

\bea \left(\frac{\dot\xi}{\xi}\right)^2=\frac{H_0^2\,e^{-2\mu\sqrt{t}}}{\xi^{3\gamma}}.\nonumber\eea By straightforward integration of this equation one finds,

\bea \xi(t)=\left(\frac{3\gamma H_0}{\mu^2}\right)^\frac{2}{3\gamma}e^{-\frac{2\mu\sqrt{t}}{3\gamma}}\left(e^{\mu\sqrt{t}}-1-\mu\sqrt{t}\right)^\frac{2}{3\gamma},\label{chi-rel}\eea so that

\bea &&\rho^{GR}_\text{rel}=\frac{M^4}{\xi^{3\gamma}}=\frac{\mu^2M^2_\text{pl}\,e^{2\mu\sqrt{t}}}{3\gamma^2\left(e^{\mu\sqrt{t}}-1-\mu\sqrt{t}\right)^2},\label{rogr}\\
&&\rho_\text{rel}=\frac{\mu^2M^2_\text{pl}}{3\gamma^2\,e^{2\mu\sqrt{t}}\left(e^{\mu\sqrt{t}}-1-\mu\sqrt{t}\right)^2}.\label{rorel}\eea The ratio of the energy densities of the relict particles according to our gauge invariant theory $\rho_\text{rel}$ and according to GR $\rho^{GR}_\text{rel}$,

\bea \frac{\rho_\text{rel}}{\rho^{GR}_\text{rel}}=e^{-4\mu\sqrt{t}},\label{ratio}\eea very quickly goes to zero. This result is independent of the behavior of the GR energy density of relic particles. 

What we have shown is that the abundance of relict particles in our theory does not represent a problem. These results are illustrated in FIG. \ref{fig3}.



\section{Discussion and conclusion}\label{sect-discu}


There are plenty of cosmological models which are designed to solve several fundamental problems in the forefront of physics. The inflationary paradigm was developed to solve the flatness, horizon and abundance of relict particles issues, among other problems at early times. The $\Lambda$CDM, quintessence and $f(R)$ models, to quote a few, are intended to explain the dark energy problem, meanwhile rock 'n' roll potential, new early dark energy, chain early dark energy and graduated dark energy models have been proposed as possible solutions to the Hubble constant tension. None of these models can take account of all of the mentioned problems at once. The question is: will be there some chance to explain several of these issues within a unified theoretical framework? Looking for a positive answer to this question has been the driving force behind this work. In this paper we have aimed at exploring the only possibility left to us by Nature for gauge symmetry to play a role in the classical description of the gravitational interactions. 


Gauge freedom can be associated with a physical picture resembling the many-worlds interpretation of quantum physics \cite{everett, dewitt, kent, barvinsky, omnes, tegmark, garriga, zurek, tegmark-nature, page}. Given that the gauge scalar $\vphi$ may be fixed at will, in equations \eqref{eom-theor}, \eqref{cons-theor} we may choose any function $\vphi({\bf x})$ we want. The result will be a specific theory associated with this choice or a gauge. Hence, each gauge represents a whole theory of gravity over WIG ($\tilde W^\text{int}_4$) space, which is characterized by a specific behavior in spacetime of several fundamental ``constants,'' the mass of the SMP particles, etc. An outstanding gauge in this theoretical framework is the so called GR gauge, which consists of a set of copies of GR theory, which are specified by the choice $\vphi=\vphi_{0i}$ ($i=1,2,...,N$), where the $\vphi_{0i}$ are different constants. In this gauge the gravitational laws look (and are) exactly the same, so that each member in the GR gauge differs from any other in the values of the fundamental constant $M^2_{\text{pl},i}$ and of the EW mass parameter $v^2_{0i}$, among others. Hence, the constant mass of given SMP particle is different in each member of the gauge. In this gauge invariant framework general relativity is just a subclass of a bigger theory. Manifest gauge symmetry is broken down once a specific gauge has been chosen. This is why GR seems to evade this symmetry. Yet, it is a residual symmetry since any gauge of \eqref{eom-theor}, \eqref{cons-theor} is related with any other gauge through the transformations Eq. \eqref{gauge-t-theor} (see the related discussion in Sec. \ref{sect-many-w}). 

Given our freedom to choose the gauge scalar $\vphi({\bf x})$, the main objection against the present approach to gauge invariance could be associated with its predictive power. Notwithstanding, on the basis of equations \eqref{eom-theor}, \eqref{cons-theor} we can make predictions as in any other theory of gravity. The only thing we have to achieve is to fully determine the gauge where we ``live'' in or our world-gauge. I. e., we need to fully determine the gauge scalar $\vphi({\bf x})$ which is consistent with the existing amount of experimental and observational evidence. This is when experiments/observations make their magic. It happens that gauge freedom can be experimentally tested in the sense that astrophysical and cosmological data sets as well as other experimental results are able to pick out, among the infinite number of equivalent gauges, our world-gauge. One example is provided by the high-redshift SN-Ia data sets \cite{riess-1998, perlmutter-1999, riess-2004, suzuki-2012}, as explained in Sec. \ref{sect-z-q}. In this case one looks for the dependence of the apparent magnitude of supernovae type Ia on the redshift. The lucky circumstance here is that, on the one hand, the apparent magnitude $m$ (do not confound with the mass parameter) is a gauge invariant quantity since it has to do with the propagation of light in spacetime: light does not interact with nonmetricity so that its propagation may be affected by spacetime curvature exclusively. On the other hand the source of light (atoms) is point dependent: the energy of atomic transitions varies from point to point in spacetime. This entails that the overall redshift is contributed both by the propagation of photons in a curved space and by the nonmetricity through spacetime variation of the atoms masses: different amounts of nonmetricity lead to different values of the overall redshift. Hence, data sets of $m(z)$ allow us to pick out a gauge which the best fit.


In the present paper we have been able just to qualitatively illustrate the possibility to determine the gauge function $\vphi({\bf x})$ by means of the check of observational data sets. It is necessary to go further and to look for new and more encompassing checks which may allow us to determine our world-gauge with more accuracy. This will be possible once we develop the theory of the cosmological perturbations that is adequate for the present gauge invariant framework. Then we will be able to make new predictions on the basis of the present theory which is based on equations \eqref{eom-theor} and \eqref{cons-theor}. 

If our theoretical framework is the one that correctly describes the gravitational phenomena in our Universe, then we have been looking for answers to the wrong questions: i) why is the cosmological constant so tiny and why its associated energy density is of the order of the present value of the dark matter energy density precisely at present?, ii) which is the nature of the dark energy?, iii) what inflates the expansion of the universe at early stages of the cosmic evolution? among others. Instead we should be wondering which is our world-gauge among the infinity of possible gauges of the gauge invariant theory of gravity.


{\bf Acknowledgments.} The author acknowledges FORDECYT-PRONACES-CONACYT for support of the present research under grant CF-MG-2558591.




\end{document}